\documentclass[12pt]{article}

\usepackage{graphicx}
\usepackage{amssymb}
\usepackage{amsmath}

\newcommand{\beq}{\begin{equation}}
\newcommand{\eeq}{\end{equation}}
\newcommand{\bea}{\begin{eqnarray}}
\newcommand{\eea}{\end{eqnarray}}
\newcommand{\ba}{\begin{array}}
\newcommand{\ea}{\end{array}}
\newcommand{\bec}{\begin{center}}
\newcommand{\eec}{\end{center}}
\newcommand{\bei}{\begin{itemize}}
\newcommand{\eei}{\end{itemize}}

\usepackage{pstricks}

\newcommand{\eq}[1]{Eq.~(\ref{#1})}

\usepackage{color}

\usepackage[body={16.5cm,22.5cm}]{geometry}

\title{\vspace{-2.6cm}
{\begin{flushright}
{\normalsize
ULB-TH/11-17\\
} 
\end{flushright}}
\vspace{2.2cm}
\bf Sub-GeV dark matter as\\ \vspace{1mm} pseudo-Goldstone from the seesaw scale\\\vspace{5mm}}

\author{Michele Frigerio$^{a,b}$, Thomas Hambye$^c$, 
Eduard Masso$^d$\footnote{E-mails: frigerio@ifae.es, thambye@ulb.ac.be}}

\date{}

\begin{document}

\maketitle

\centerline{$^a$ {\it \footnotesize CNRS, Laboratoire Charles Coulomb, UMR 5221, F-34095 Montpellier, FRANCE} }
\centerline{\it   \footnotesize  Universit\'e Montpellier 2, Laboratoire Charles Coulomb, UMR 5221, F-34095 Montpellier, FRANCE}
\centerline{$^b$ \it    \footnotesize  Institut de F\'isica d'Altes Energies, Universitat Aut\`onoma de Barcelona, E-08193 Bellaterra, SPAIN}
\centerline{$^c$ \it   \footnotesize  Service de Physique Th\'eorique, Universit\'e Libre de Bruxelles, 1050 Bruxelles,
BELGIUM}
\centerline{$^d$ \it   \footnotesize  Grup de F{\' \i}sica Te{\`o}rica, Departament de F{\' \i}sica,
Universitat Aut{\`o}noma de Barcelona, E-08193 Bellaterra, SPAIN}

\vspace{4cm}

\begin{abstract}
Pseudo Nambu-Goldstone bosons (pNGBs) are naturally light spin-zero particles, which can be interesting dark matter (DM) candidates.
We study the phenomenology of a pNGB $\theta$ associated with an approximate symmetry of the neutrino seesaw sector.
A small coupling of $\theta$ to the Higgs boson is induced radiatively by the neutrino Yukawa couplings.
By virtue of this Higgs portal interaction
(i) the pNGB acquires a mass $m_\theta$ proportional to the electroweak scale, and
(ii) the observed DM relic density can be generated by the freeze-in of $\theta$-particles
with mass $m_\theta \simeq 3$ MeV. Alternatively, the coupling of $\theta$ to heavy sterile neutrinos can 
account for the DM relic density, in the window $1$ keV $\lesssim m_\theta \lesssim 3$ MeV.
 The decays of $\theta$ into light fermions are suppressed by the seesaw scale, making such pNGB sufficiently stable to play the role of DM.
\end{abstract}
\vspace{5mm}

\newpage
\section{Introduction}

The astrophysical and cosmological evidence for dark matter (DM) 
requires that a new particle is introduced,
with constrained mass and interactions. These interactions must in particular maintain the DM candidate stable on cosmological time scales
and generate the observed DM relic abundance.
Many candidates can satisfy all constraints. Much fewer, though, provide a convincing explanation for the required values of the DM parameters.
As examples, 
in most models (i) an ad-hoc 
symmetry is postulated to insure stability, and/or (ii) the DM mass is assumed to be of order (or below) the electroweak (EW) scale as 
demanded by the Weakly Interacting Massive Particle paradigm, and/or (iii) new interactions 
with strength similar to the EW ones,
not motivated by other means, are just postulated. 
A scenario which is potentially predictive, and could provide motivations for some of these assumptions, consists in identifying the DM candidate with the pseudo-Nambu-Goldstone boson (pNGB) of a spontaneously broken global symmetry.

An interesting feature 
of such scenario 
is that
the DM couplings to the Standard Model (SM) fields, which could render the DM candidate unstable on cosmological times, 
 are suppressed by powers of the spontaneous symmetry breaking (SSB) scale. 
As a result, provided this scale is large, the decay width of the particle can 
be sufficiently small.
A well-known example is the axion, 
the pNGB associated with the Peccei-Quinn $U(1)$ symmetry, broken spontaneously at the scale $f_{PQ}$. 
The axion is a viable DM candidate when $f_{PQ}$ lies in the interval $\sim (10^{12}-10^{14})$
GeV.
Another possibility is to consider SSB at the seesaw scale, which should be  large to account for the smallness of the neutrino masses.
One global symmetry related to the seesaw is $U(1)_{B-L}$.  In the case that it is spontaneously broken by an EW singlet, coupled to 
heavy sterile (right-handed) neutrinos, the corresponding
NGB is known as the singlet Majoron \cite{CMP}. 
Models with a pseudo-Majoron as DM have been studied \cite{ABST}-\cite{GMS}. 
More generally, we will consider  
any (family-dependent) global symmetry whose spontaneous breaking contributes to sterile neutrino masses. 
Symmetries associated with the seesaw sector have the advantage that the pNGB interactions are anticipated to exist anyway to account for neutrino masses.

Another remarkable property of the pNGB scenario, relevant for the DM phenomenology,
is that the mass and the scalar potential interactions of a pNGB vanish in the limit of exact symmetry.
Therefore, when the symmetry is explicitly broken by a unique (or dominant) term, the scalar interactions and the DM mass are necessarily related, 
and, if the relic density is essentially determined by such interactions, this implies a one-to-one correspondence  
between the DM relic density and mass.  
For instance we will consider a framework where the source of explicit breaking generates radiatively a `Higgs portal' interaction of the form
$\lambda \theta^2 H^\dagger H/2$, 
with $\theta$ the pNGB particle and $H$ the SM Higgs doublet. This coupling generates a mass $m^2_\theta =
\lambda v^2$ with $v=174$ GeV and, at the same time,  a relic density from the 
$\theta - H$ interactions.
We will show how, given the known thermal distribution of SM particles in the early universe thermal bath, 
the Higgs portal can generate a DM relic density through the freeze-in mechanism (see e.g. Ref.~\cite{Hall:2009bx}). 
The observed relic density is obtained for a unique possible value of the DM mass, which turns out to be $m_\theta \simeq 2.8$~MeV 
for a Higgs boson mass of 120~GeV.
There is also a second possibility to generate the correct relic density, by the freeze-in of the interaction of $\theta$
with the sterile neutrinos.
These production mechanisms differ from those previously considered for a pNGB candidate for DM.

A third interesting feature of the pNGB setup is that the DM mass is related to the scale of explicit symmetry breaking, 
which can be identified with a physical mass scale already present in the theory,
analogously to the QCD scale for the axion.
For instance, in the scenario we will consider below, the neutrino Dirac Yukawa couplings provide the source of explicit breaking of the global symmetry.
Therefore, the DM mass is proportional to the EW scale, $m^2_\theta = \lambda v^2$,
and this  provides a justification for the presence of a scalar DM particle at or below the EW scale. 
Moreover, the coupling $\lambda$ 
can be computed as a function of the seesaw couplings. 
In turn, we will show that in this framework the DM mass can be protected from quadratically divergent corrections.
The mechanism we invoke to remove these corrections is a variation of the one proposed long ago by Hill and Ross 
for very light pNGBs 
with both scalar and pseudo-scalar couplings \cite{HR}.
We will consider a $U(1)_X$ symmetry such that its explicit breaking requires several Yukawa couplings at the same time,
thus lowering the degree of divergence of the contributions to the pNGB effective potential.
This mechanism has similarities with the collective breaking mechanism introduced to protect the EW scale in 
little Higgs models, where the SM Higgs is the pNGB (see Ref.~\cite{LH} for a review).

All in all, we will show that the seesaw interactions (i) are associated with global symmetries broken spontaneously at the heavy neutrino mass scale,
whose largeness guarantees the stability of the pNGB DM candidate, and (ii) have a source of explicit symmetry breaking built-in, and therefore induce the mass and scalar interactions of the DM. Moreover, they can lead to (iii) a one-to-one correspondence between the DM mass and 
the DM relic density, with (iv) a justification of the presence of DM at low scale, and with (v) a DM mass  protected from large radiative corrections.

The paper is organized as follows.
In section 2 we display the relevant effective interactions of our pNGB candidate for DM.
In section 3 we show how a pNGB relic density of the order of the observed DM density can be obtained through freeze-in.
In section 4 we present a class of seesaw models where the required pNGB interactions are generated naturally.
In section 5 we compute the pNGB couplings to SM particles and analyze the corresponding constraints on the DM stability.
We conclude in section 6.

\section{Effective interactions of the pNGB}

In this section 
we introduce the effective lagrangian of a pNGB $\theta$, associated with
a global symmetry of the neutrino sector, broken spontaneously at the seesaw scale $f$. 
We postpone to section \ref{appro} the detailed description of a class of models where such a light pNGB may emerge naturally.
For the present purposes, it suffices to specify the effective interactions of $\theta$ at low energy (below $f$).

In the class of models under consideration, $\theta$ couples to a sterile 
neutrino $\nu^c$ as follows:
\begin{equation}
\label{LN}
{\cal L}_N = -\frac{gf}{2\sqrt{2}} \nu^c\nu^c e^{i\theta/f} + {\rm h.c.} = -\frac{m_N}{2} \overline{N}N 
+\frac{ig}{2\sqrt{2}} \theta \overline{N}\gamma_5 N  +\frac{g}{4\sqrt{2}f} \theta^2 \overline{N} N + \dots ~,
\end{equation}
where $g$ is a Yukawa coupling and the 4-component Majorana neutrino is defined as usual by $N\equiv(\nu^c~\nu^{c\dag})^T$.
For simplicity, we  consider only one sterile neutrino, whose mass $m_N = gf/\sqrt{2}$ is generated by SSB.
 In this case the lagrangian shown in \eq{LN} is the same as for the singlet Majoron model \cite{CMP}, where the spontaneously broken
global symmetry is lepton number.
However, we will show in section \ref{appro} that in realistic cases with two (or more) sterile neutrinos the relevant symmetry is not the ordinary lepton number, rather each sterile neutrino carries a different charge, 
and consequently the pNGB $\theta$ cannot be identified with the Majoron.
We remark that the $\theta-N$ interaction is just an example of a NGB interaction with a heavy fermion charged under the associated global
symmetry. As a consequence, the  implications for dark matter phenomenology, discussed in the following sections, hold qualitatively 
in a more general set of theories, with heavy fermions other than the sterile neutrinos.

A non-vanishing potential for the NGB is generated when one introduces a source of explicit breaking of the global symmetry. 
In the models we will consider, this source is given by a certain set of neutrino Dirac Yukawa couplings, 
so that the $\theta$ effective potential necessarily involves the SM Higgs doublet $H$. The relevant, radiatively induced coupling is
given by 
\begin{equation}
\label{LH}
{\cal L}_H = -\frac{\lambda}{2}   \theta^2   H^\dagger H  = 
- \frac{m_\theta^2}{2} \theta^2 - \frac{\lambda v}{\sqrt{2}} \theta^2 h 
- \frac{\lambda}{4} \theta^2 h^2 ~,
\end{equation}
where $v\simeq 174$ GeV is the Higgs vacuum expectation value (vev), $h$ is the physical Higgs boson 
and we adopted the unitary gauge. 
Note that the mass of the pNGB $\theta$ is proportional to the EW vev, $m_\theta^2= \lambda v^2 $. 
This relation is
sometimes referred to as the `conformal' limit, since $\theta$ has no bare mass term, rather 
the Higgs portal interaction $\lambda$ generates both $m_\theta$ and the $\theta-h$ couplings. We remark that the phenomenology (e.g.~the DM mass, its relic density, etc.) induced by the Higgs portal in \eq{LH}
does not depend on the details of the underlying physics  which generates radiatively $\lambda$.

In the specific seesaw models that we will build, $\lambda$ is generated by logarithmically divergent neutrino loops, involving both the 
$\theta-N$ coupling $g$ as well as neutrino Dirac Yukawa couplings, denoted generically with $y$. 
It turns out that the relation between the mass of $\theta$ and the neutrino parameters can be written schematically as
\beq
m^2_\theta = \lambda v^2 \simeq g^2 y^2 v^2 
\ \frac{\log(\Lambda^2/m_N^2)}{8\pi^2} ~,
\label{mesti}\eeq 
where $\Lambda$ is some cutoff scale at or above $f$. The light neutrino mass scale is given by the standard seesaw relation, $m_\nu=y^2v^2/m_N$.
It is useful to use \eq{mesti} to express the coupling $g$ as a function of the relevant energy scales in the theory:
\beq
g = 10^{-3}\left(\frac{m_\theta}{{\rm MeV}}\right) \left(\frac{{\rm eV}}{m_\nu}\right)^{1/2}
\left(\frac{10^9{\rm GeV}}{m_N}\right)^{1/2} \left(\frac{8\pi^2}{\log(\Lambda^2/m_N^2)}\right)^{1/2}~.
\label{lagi}
\eeq
We will take the factor $k\equiv \log(\Lambda^2/m_N^2)/(8\pi^2)$
to be of order one. Then, for a fixed value of $m_\nu$,
$m_N$ and also $f=\sqrt{2}m_N/g$ are determined as a function of $m_\theta$ and $g$, which 
are basically the only two free parameters.
We require the scale $f$ to be smaller than the Planck scale, $M_P=1.22 \times 10^{19}$ GeV.
In addition, we assume for simplicity that $f$ is larger than one TeV, 
in order for the effective theory to contain only the SM particles, the pNGB and the sterile neutrino,
while the degrees of freedom involved in the SSB are decoupled.
These two requirements exclude
the green shaded regions in the $(m_\theta - g)$ plane shown in Fig.~\ref{fig1}.
The dotted lines in Fig.~\ref{fig1} correspond to constant values of the sterile neutrino mass,
$m_N=m_\theta^2/(g^2 m_\nu k)
=10^2,10^6,10^{10},10^{14}$ GeV, from top to bottom.

\section{Relic density from freeze-in of the pNGB \label{relic}}

We now study the role of the interactions in Eqs.~(\ref{LN}) and (\ref{LH})
for the production of a $\theta$ relic density, which can play the role of DM.
In this section, we implicitly assume that $\theta$ is stable on cosmological time scales. The region of parameters where this stability is achieved
 will be analyzed in detail in section \ref{intau}.
There are several mechanisms to produce the DM relic density.
We begin by briefly recalling  these various possibilities and
confront them 
with the parameter space of our scenario.

A well-known way to produce a cosmological density of a DM particle species is the usual freeze-out: it requires an annihilation cross section large enough to overcome the expansion of the universe and thermalize the DM particle. The interactions of $\theta$ come from the terms in 
\eq{LN}, which we refer to as the `sterile neutrino portal', and the terms in \eq{LH}, that is, the `Higgs portal'. These interactions may or may not lead to the thermalization of $\theta$ in the early 
universe, depending on whether the  interaction rate $\Gamma$ gets larger than the Hubble rate $H$. 
If it does, later on $\theta$ necessarily decouples from the plasma and, if stable, acts as DM with a certain 
energy density  $\rho_\theta$, which depends in general on its mass and annihilation cross section. 

\begin{figure}[t!]
\begin{center}
\includegraphics[width=13cm]{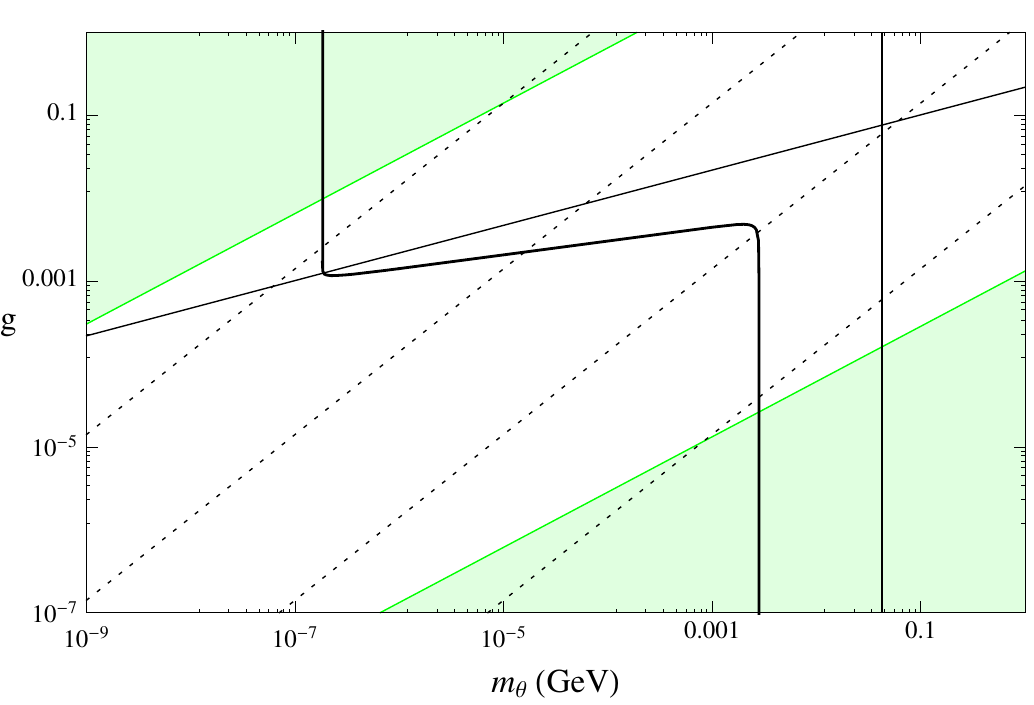}
\caption{\small The pNGB coupling to the sterile neutrino, $g$, versus the pNGB mass, $m_\theta$, in GeV.
We fixed $m_\nu=0.05$ eV and $k=1$.
The upper (lower) green shaded region is excluded by requiring the  SSB scale $f$ to be larger than
one TeV (smaller than the Planck scale). The four dotted lines correspond to constant values of the sterile neutrino mass:
from top to bottom, $m_N=10^2$, $10^6$, $10^{10}$, $10^{14}$ GeV.
The thin diagonal (vertical) line indicates the lower value of $g$ ($m_\theta$) that leads to a thermalization of $\theta$.
The thick line corresponds to a $\theta$ relic density equal to the observed DM relic density, as follows from the numerical solution
of the Boltzmann equation  (for $m_h=120$ GeV).
Since $\theta$ is produced by freeze-in, its relic density grows with its couplings $g$ and $\lambda = m_\theta^2/v^2$,
therefore the region below and to the left  (above and to the right) of the thick
line corresponds to $\Omega_\theta <  \Omega_{DM} (> \Omega_{DM})$.
}
\label{fig1}
\end{center}
\end{figure}

Before determining
$\rho_\theta$ numerically, 
it is instructive to identify the thermalization region in first  approximation, 
by evaluating the $\theta$ interaction rate $\Gamma$ and 
requiring $\Gamma > H$. 
We start with the Higgs portal, considering for simplicity the 
$h\rightarrow \theta \theta$ decay only, because in our scenario decays turn out to dominate with respect to 
scattering processes (see Fig.~\ref{fig0} below). The decay rate is
\begin{equation}
\Gamma(h\rightarrow \theta \theta)=\frac{1}{16 \pi} \lambda^2 \frac{v^2}{m_h}\sqrt{1-\frac{4m_\theta^2}{m_h^2}} ~.
\label{Gammah}
\end{equation} 
This should be compared to the Hubble parameter in the radiation epoch, which is given by 
$H=1.66 \sqrt {g_*^\rho}\, T^2/M_P$,
with $g_*^\rho=106.75$ the degrees of freedom corresponding to the SM content
(we shall assume such value in the rest of the paper).
Taking $T\simeq m_h$ to be the relevant temperature, we find that, for a coupling
\begin{equation}
\lambda \gtrsim 6 \times 10^{-8} \left( \frac{m_h}{120\ {\rm GeV} } \right)^{3/2}~, 
\label{eq6}\end{equation}
the decays and inverse-decays
are able to produce a thermal population of $\theta$-particles. In other words, given the $m^2_\theta=\lambda v^2$ relation, one needs $m_\theta > 44$~MeV, which translates into the thin vertical line in Fig.~\ref{fig1}.\footnote{Notice that \eq{eq6} is the condition for DM to thermalize with 
the SM sector at  $T \simeq m_h$ (where the rates reach the maximum, see Fig.~\ref{fig0} below). This condition is different from the one usually considered for the freeze-out scenario, where one requires that the scattering processes are in thermal equilibrium at $T \lesssim m_\theta$. 
The latter condition leads to a value of $\lambda$ about two orders of magnitude larger than the one in \eq{eq6}.}

Similarly for the interaction with the sterile neutrino $N$,
the process $NN \rightarrow \theta \theta$ has a rate
\begin{equation}
\Gamma = \langle \sigma v_{rel}  \rangle n_N \simeq  \frac{g^4}{256 \pi} \frac{n_N}{m_N^2}
\simeq 2 \times10^{-4} \  g^4 m_N ~,
\end{equation}
with $n_N$ the sterile neutrino number density, and we used $T\simeq m_N$. This leads to thermal production of $\theta$-particles provided 
\begin{equation}
g\gtrsim  10^{-2} \left(\frac{\rm eV}{m_\nu k}\right)^{1/6} \left(\frac{m_\theta}{\rm MeV}\right)^{1/3}~,
\label{gth}\end{equation}
where we made use of \eq{lagi}. This gives the thin diagonal line in Fig.~\ref{fig1}.
We do not discuss  the effects of the couplings of $\theta$ to light fermions, because in the models under consideration
they turn out to be far too small to thermalize $\theta$ (see section \ref{intau}).

Outside the thermalization region, where $\Gamma$ is smaller than $H$, one may still have a final $\Omega_\theta$ matching $\Omega_{DM}$
(as usual we define $\Omega_i \equiv \rho_i/\rho_{crit}$, i.e. the ratio between the energy density $\rho_i$ and the critical density of the universe).
The idea is that in the early universe a population of $\theta$-particles 
is produced through pair production scattering processes, but with a density smaller than the thermal one. 
For appropriate values of $m_\theta$ one may still reach a relic density equal to $\Omega_{DM}$. 
This process, called freeze-in, actually happens in our scenario as we shall describe below in detail
(for generic properties of the freeze-in mechanism see Ref.~\cite{Hall:2009bx}).

In the rest of this section we first introduce the ingredients to make a numerical calculation of the $\theta$ relic density,
and we then apply them to the freeze-out and freeze-in processes. 
Afterwards we also discuss some non-thermal production mechanisms and end up with a summary of our results.

\subsection{Boltzmann equation and reaction densities}

From the Higgs portal,  
$\theta$-particles are created 
from the following 
scattering processes:\footnote{In Eqs.~(\ref{sigmahiggsh})-(\ref{NNtt}) we included the usual initial-state spin and color average factor, as
well as the final state identical particle factor $1/2$. 
In $\sigma(h h \rightarrow \theta \theta)$ we neglected subleading terms of order $\lambda^3$ and $\lambda^4$ 
(which are displayed e.g. in Ref.~\cite{GW}).}
\begin{eqnarray}
\sigma(h h \rightarrow \theta \theta)  &\simeq& \frac{\lambda^2}{32 \pi s}\left(\frac{s-4 m^2_\theta}{s-4 m^2_h}\right)^{1/2}  
\left(\frac{s+2 m_h^2}{s- m_h^2}\right)^2\,, 
\label{sigmahiggsh}\\
\sigma(W W \rightarrow \theta \theta)  &=&\frac{1}{9}\frac{\lambda^2}{32 \pi s}  \left(\frac{s-4 m^2_\theta}{s-4 m^2_W}\right)^{1/2} \
\frac{s^2-4 s m^2_W+12 m_W^4}{(s-m_h^2)^2+m_h^2 \Gamma_h^2}   \,,\label{sigmahiggsW}\\
\sigma(Z Z \rightarrow \theta \theta)  &=&\frac{1}{9}\frac{\lambda^2}{32 \pi s} \left(\frac{s-4 m^2_\theta}{s-4 m^2_Z}\right)^{1/2} \
\frac{s^2-4 s m^2_Z+12 m_Z^4}{(s-m_h^2)^2+m_h^2 \Gamma_h^2}    \,,\label{sigmahiggsZ}\\
\sigma(f \bar{f}  \rightarrow \theta \theta)  &=&\frac{1}{4} \frac{1}{n_c}\,   \frac{\lambda^2}{16\pi  s} 
 \frac{m_f^2(s - 4 m_f^2)^{1/2} (s-4 m_\theta^2)^{1/2}}{(s-m_h^2)^2+m_h^2 \Gamma_h^2}  \,.\label{sigmahiggsf}
\end{eqnarray}
where $\Gamma_h$ is the total Higgs decay width and $n_c$ is the number of colours of the fermion $f$. 

Similarly, $\theta$-particles can be created through the scattering with a sterile neutrino,
\begin{equation}
\sigma(NN\rightarrow \theta\theta) \simeq \frac{g^4}{256 \pi} \, \frac{1}{\beta_N^2 \, s} \left( \beta_N \frac{s}{m_N^2} + 
2 \log \frac{1 - \beta_N}{1 + \beta_N} \right)\,,
\label{NNtt}
\end{equation} 
where we neglected $m_\theta$ in front of $m_N$ and $\sqrt{s}$ and we defined $\beta_N= (1 - 4 m_N^2/s)^{1/2}$.

In the following it will be convenient to consider separately the effect of the Higgs decay, 
Eq.~(\ref{Gammah}), which means that,  in order to avoid double counting, 
the on-shell part of the scattering processes must be subtracted (whenever there is one).
In the narrow width approximation this means to perform in Eqs.~(\ref{sigmahiggsW})-(\ref{sigmahiggsf}) 
the substitution
\begin{equation}
\frac{m_h \Gamma_h}{(s-m_h^2)^2+m_h^2 \Gamma_h^2}\quad \rightarrow \quad \frac{m_h \Gamma_h}{(s-m_h^2)^2+m_h^2 \Gamma_h^2}
- \pi \delta(s-m^2_h)\theta(s-4m^2_i)~,
\label{substraction}
\end{equation}
for $i=W,Z,f$. 
In the following, all scattering quantities we will consider are meant to be the subtracted ones, using Eq.~(\ref{substraction}).
To calculate the relic density that one obtains from these processes, either through freeze-in or freeze-out, 
one has to integrate the Boltzmann equation,
\begin{equation}
zH(z)s(z) Y'_{\theta}(z) = \left[1-\left(\frac{Y_{\theta}(z)}{Y_{\theta}^{\rm eq}(z)}\right)^2 \right] (\gamma_{D_h}(z)+\gamma_{annih}(z))\,,
\label{Boltz}
\end{equation}
with $z\equiv m_h/T$ conventionally taken as the evolution parameter. Here $H(z)$ is the Hubble parameter,
$s(z)$ the entropy density,
and $Y_\theta\equiv n_\theta/s$ with $n_\theta$ the number density. The reaction density $\gamma_{D_h}$ contains the effect of the Higgs boson decay, 
while $\gamma_{annih}$ is the sum of the (subtracted) reaction densities of the scattering processes in Eqs.~(\ref{sigmahiggsh})-(\ref{NNtt}). The reaction densities
are given by
\begin{eqnarray}
\gamma_{D_h}&=&   N_\theta  \int d\bar{p}_h f_h^{eq} 
\iint d\bar{p}_1 d\bar{p}_2 (2\pi)^4 \delta^4(p_h-p_1-p_2) |{\cal M}|^2 \nonumber\\
&=& N_\theta \, n_h^{eq} \frac{K_1(z)}{K_2(z)}\, \Gamma(h\rightarrow \theta \theta)\,,\\
\gamma(a\,b\leftrightarrow 1\,2) &=& N_\theta  \iint d\bar{p}_a d\bar{p}_b 
f_a^{eq}f_b^{eq} \iint d\bar{p}_1 d\bar{p}_2 (2\pi)^4 \delta^4(p_a+p_b-p_1-p_2) |{\cal M}|^2 \nonumber\\
&=& N_\theta \, \frac{T}{64~\pi^4} \int_{s_{min}}^{\infty} ds ~\sqrt{s}~\hat{\sigma}(s)~K_1\left(\frac{\sqrt{s}}{T} \right)\,,
\label{ScatRates}
\end{eqnarray}
involving the Bessel functions $K_{1,2}$. We have defined $d \bar{p} \equiv 
d^3 p/((2 \pi)^3 2 E)$. Here $N_\theta=2$ is the number of $\theta$ particles produced per decay or annihilation, 
$f_i^{eq}=(e^{E_i/T}\pm 1)^{-1}\simeq e^{-E_i/T}$ is the Maxwell-Boltzmann energy distribution,  
$|{\cal M}|^2$ is the amplitude squared summed over initial and final spins (with no averaging),
$s_{min}=\hbox{max}[(m_a+m_b)^2,(m_1+m_2)^2]$, and  the reduced cross section is defined by
\beq
\hat{\sigma}(a\,b\leftrightarrow 1\,2) = \frac{g_ag_b}{c_{ab}}\ \frac{2[
(s-m_a^2-m_b^2)^2 -4 m^2_a m^2_b]}{s} \ \sigma(a\,b\rightarrow 1\,2) \,,
\eeq
with $\sigma$ the particle physics cross section of Eqs.~(\ref{sigmahiggsh})-(\ref{NNtt}), $g_{a,b}$ the number of degrees of freedom of the particles 
$a,b$  and $c_{ab}$ a combinatorial factor equal to 2 (1) if $a$ and $b$ are identical (different).

\subsection{Freeze-out \label{fo}}

As it is well-known, if $\theta$ freezes out relativistically, that is at $T\gtrsim m_\theta$, its relic density is independent of the annihilation cross section,
because it decouples when the thermal number density, $Y_\theta^{eq}\equiv n_\theta^{eq}/s$ is still independent of the temperature. 
The formula for the relative density $\Omega_\theta$ 
is simply
\begin{equation}
\Omega_\theta h_0^2 \simeq 78\,  \frac{1}{g_*^s} \, \frac{m_\theta}{\rm keV} ~,
\end{equation}
with $g_*^s$ the number of degrees of freedom contributing to entropy at $\theta$ decoupling.
This should match the value
\begin{equation}
\label{Omega_DM}
\Omega_{DM} h_0^2 = 0.11 \pm 0.01 ~.
\end{equation}
In these equations, $h_0$ is the reduced Hubble constant $h_0=H_0/(100$ Mpc km s$^{-1}) \simeq 0.70$.
As a result, the observed relic density can be obtained only for one DM mass value:
$m_\theta\simeq 0.15$~keV, where we took $g_*^s =106.75$, valid for a decoupling temperature above 100 GeV.
Such a light $\theta$ can be only thermalized by the interaction with sterile neutrinos, which decouple
at a temperature just below $m_N$, leading to relativistic $\theta$'s.
This result corresponds to the left vertical branch of the thick curve in Fig.~\ref{fig1},
which is obtained by integrating numerically the Boltzmann equation.

If instead $\theta$ is heavier and thermalizes, to have a small enough relic density requires that it decouples non-relativistically. This can occur through the Higgs portal interaction.
In this case the relic density essentially only depends on the annihilation cross section which must take the value
$\langle \sigma v_{rel}\rangle \simeq 10^{-26}$~cm$^3$s$^{-1}$.
To get a sufficiently small relic density,
one needs to go to much higher values of $m_\theta$, where the coupling $\lambda=m^2_\theta/v^2$
becomes large enough. 
The observed relic density is obtained for a unique value of $m_\theta$, because of the relation between $m_\theta$ and $\lambda$,
which is a consequence of the pNGB nature of $\theta$.
The freeze-out through the Higgs portal was considered  in Ref.~\cite{Farina:2009ez}, assuming
this restrictive relation, and it was found 
that the DM relic density is obtained with 
$m_{\theta}=50-70$~GeV (for $m_h=120-180$~GeV).

However, as we will see in section \ref{intau}, the freeze-out values, $m_\theta=0.15$ keV as well as $m_\theta=50-70$ GeV,
require values of parameters which are excluded
by the instability of $\theta$ on cosmological time scales.
Therefore in our scenario both relativistic and non-relativistic freeze-out is not a viable way to produce DM. Instead,
the freeze-in process turns out to be efficient in an allowed region of the parameter space, as we now discuss.

\subsection{Freeze-in \label{FI}}

In the freeze-in process $\theta$ is produced by
the annihilation or decay of a heavier particle $X$, with rates small enough not to thermalize, leading 
to a less-than-thermal DM population, which reaches a plateau at $T\sim m_X$ because at smaller $T$ the number of $X$-particles is Boltzmann suppressed. 
If the $X$-particles are the SM particles, as it is the case through the Higgs portal, their number densities are known as a function of the temperature 
(simply given by their thermal distribution down to a temperature well below their mass). As a consequence, the created relic density depends only on the magnitude of the portal.\footnote{For production of DM through the Higgs portal along the freeze-in scenario, see also Ref.~\cite{Yaguna:2011qn}.}

Let us first consider a value of the coupling $g$ sufficiently small  to neglect $\theta$ production at the seesaw scale
(we will see that this is the case for $g\lesssim 10^{-3}-10^{-4}$, depending on $m_\theta$). 
In this case only the Higgs portal interactions can account for the observed relic density. 
As said above, this mechanism is extremely predictive since, given the `conformal' relation $m^2_\theta=\lambda v^2$, the rates and hence the relic density depend only on the 
parameter $m_\theta$.

Integrating numerically Eq.~(\ref{Boltz}), it turns out that the observed relic density 
in \eq{Omega_DM} is obtained for 
\begin{equation}
m_{\theta}= (2.76 - 2.86)~\hbox{MeV} ~,
\label{mdmconformal}
\end{equation}
where we took $m_h=120$~GeV for definiteness.
This result corresponds to the right vertical branch of the thick curve in Fig.~\ref{fig1}.

\begin{figure}[t]
\begin{center}
\includegraphics[width=8cm]{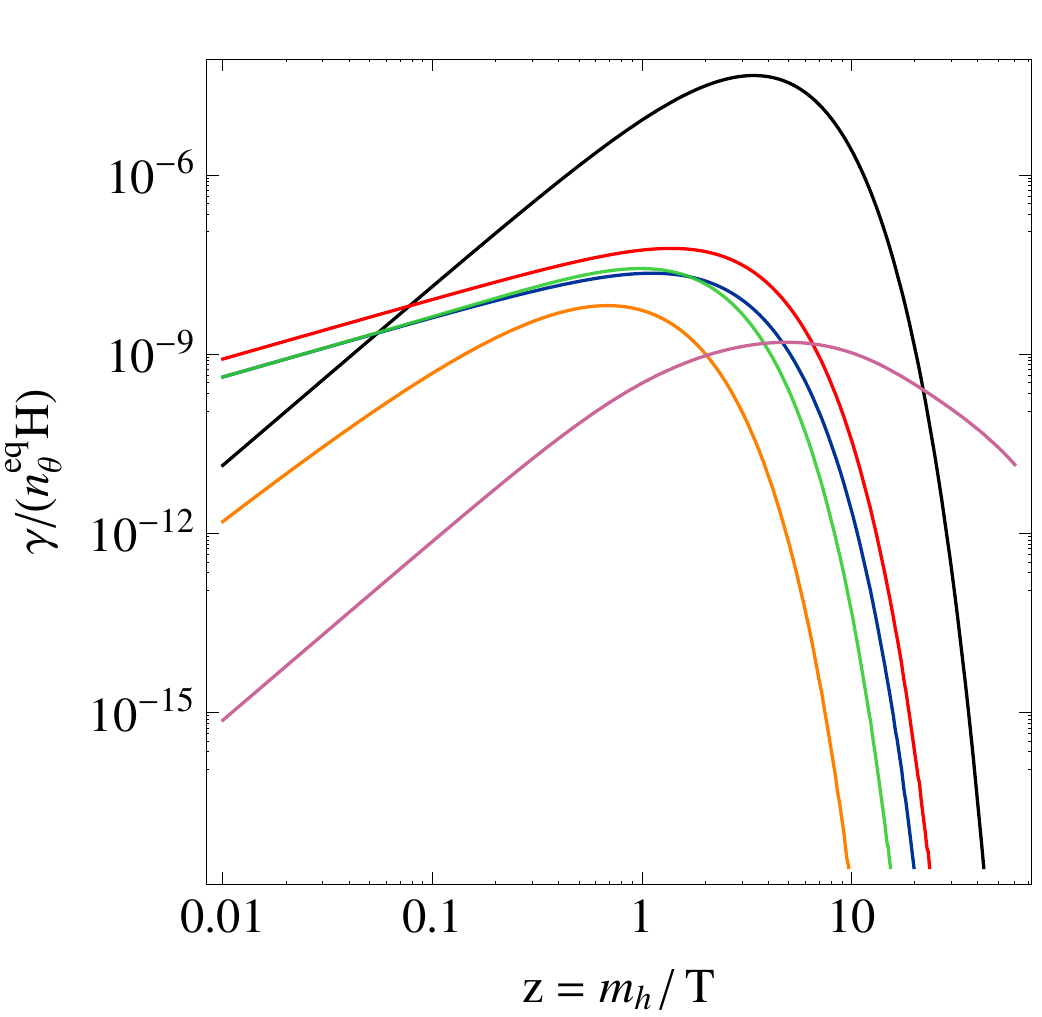}
\caption{\small
The decay thermalization rate, $\gamma_{Dh}/(n_\theta^{eq} H)$ (black), compared to the various scattering thermalization rates, 
$\gamma^a_{annih}/(n_\theta^{eq} H)$ for $a=W,Z,h,t,b$ (in red, blue, green, orange and purple respectively),
as a function of $z=m_h/T$ and for $m_h=120$~GeV, $m_\theta=2.8$~MeV. 
For freeze-in these rates remain always well below one, which is the thermalization value.
} 
\label{fig0}
\end{center}
\end{figure}

An important difference between freeze-in and freeze-out is that, while the latter is generally dominated by the off-shell annihilation, the former is naturally dominated by the on-shell annihilation, i.e.~by the decay of the mediator involved in the annihilation. For a freeze-out, unless the mediator has a mass $\sim 2 \,m_{DM}$, the decay has a small effect because, when the off-shell annihilation goes out-of equilibrium,
it has already decoupled (being more Boltzmann suppressed). 
For a freeze-in instead, the DM production occurs mostly at $T$ of order the mediator mass, or slightly below, where the Boltzmann suppression
is still mild. As a result the decay, which involves less (small) couplings, is naturally dominant, $\gamma_{D_h} > \gamma_{annih}$.
This is shown in Fig.~\ref{fig0} where the various reaction densities are plotted for $m_h=120$~GeV and $m_\theta=2.8$~MeV.
In the approximation where only the decay is included, the Boltzmann equation can be integrated analytically, giving \cite{Hall:2009bx}
\begin{equation}
Y_{\theta} \simeq N_\theta  \frac{135 \,g_{h}}{8 \pi^3 (1.66) g_*^s\sqrt{g_*^\rho}}\frac{M_P \Gamma(h\rightarrow \theta \theta)}{m_h^2}\,,
\label{Yanaly}
\end{equation}
where $g_h=1$.  
This analytic result turns out to differ from the numerical result by less than 1\%.

To understand Eq.~(\ref{Yanaly}) let us first discuss why the freeze-in is infrared dominated,
with most of the $\theta$ production occurring 
at $T\lesssim m_h$ (just before $n_h^{eq}$ gets too much Boltzmann suppressed). 
The $\theta$ number density produced at temperatute $T$ is essentially given by the number of decays per unit time occurring at this 
 temperature, 
$n_h^{eq} (m_h/T)\Gamma(h\rightarrow \theta\theta)$,
 times the number of $\theta$-particles produced per decay, $N_\theta$,
times the Hubble time, $1/H$. Therefore, taking into account thermal averaging 
properly, one has 
$Y_\theta \simeq \gamma_{Dh}/(Hs)$, which is nothing but the rate of thermalization $\gamma_{Dh}/(n_\theta^{eq} H)$ shown in Fig.~\ref{fig0}, 
times the relativistic value of $Y_\theta^{eq}$.
Thus one obtains 
$Y_\theta 
\propto M_{P} m_h \Gamma(h \rightarrow \theta \theta)/T^3$, which increases as $T$ decreases. 

Hence one finds 
$Y_{\theta}= c [N_\theta n_h^{eq} \Gamma(h\rightarrow \theta\theta)/(H s)]_{T=m_h}$, 
with $c$ a numerical factor which turns out to be equal to $(3\pi/2)/K_2(1) \simeq 2.9$. It is larger than one because, as Fig.~\ref{fig0} shows, the maximum value of the decay reaction density occurs 
at $T\sim m_h/3.5$ rather than at $T\sim m_h$.\footnote{Actually if one takes the value of 
$\gamma_{Dh}/(Hs)$
at its maximum value, i.e.~at $T\sim m_h/3.5$, one obtains 
a number density equal to $70\%$ of the exact result. Therefore, the 
production of a particle through freeze-in of a decay is essentially given by the maximum value of the corresponding thermalization rate.
Note also that if one takes instead $N_\theta n_h^{eq} \Gamma(h\rightarrow \theta\theta)/(H s)$ at its maximum value, which occurs for $T\sim m_h/3.0$, 
it turns out that one gets an even better approximation, as this overestimates the exact result by only $3\%$.}
With this in mind, and taking into account the dependence $\Gamma(h\rightarrow \theta \theta) \propto \lambda^2/m_h \propto m_\theta^4/m_h$,
 the DM relic density from the Higgs decay scales as $m_\theta^5/m_h^3$, so that the value of $m_\theta$ needed varies very little with 
the exact value of the relic density, and the larger the Higgs mass is, the larger $m_\theta$ has to be.
For example for $m_h=140,180,300$~GeV one gets the observed relic density for $m_\theta=3.0,3.6,4.9$~MeV instead of $2.8$~MeV in Eq.~(\ref{mdmconformal}).

It is interesting to discuss what is the contribution of each scattering channel separately, even though they have a small effect with respect to the decay.
As can be seen from Fig.~\ref{fig0}, among the scattering channels, the $W$ one gives a contribution slightly larger than the $Z$ and $h$ ones, and much larger than the $b$ and $t$ ones.
Numerically, for $m_h=120$~GeV, the $W,\,Z,\,h,\,b,\,t$ scatterings contribute 
to the {total} number of $\theta$-particles produced
in the proportions $1:0.4:0.4:4\times 10^{-3}:10^{-1}$
 respectively, while the $W$ scattering alone gives a contribution 500 times smaller than the decay channel.

Similarly to the decay, at a temperature $T\gtrsim m_a$, a scattering process $a a \rightarrow \theta \theta$ gives a number density 
$Y_{\theta}^a \simeq \gamma^a_{annih}/(H s)$, 
which is infrared dominated too.
Again, this is nothing but the thermalization rate $\gamma^a_{annih}/(n_\theta^{eq} H)$, displayed in Fig.~\ref{fig0},
up to the multiplicative constant $Y_\theta^{eq}$. One gets  $Y_{\theta}^a = c_a   [\gamma^a_{annih}/(H s)]_{T=m_a}$ with $c_a$ 
a numerical factor which, unlike for the decay, does not take a unique value. It depends on $m_\theta$ as well as on the $s$ dependence of the scattering cross section (which fixes the position of the peak of the reaction densities in Fig.~\ref{fig0}). For the $W$, $Z$ and $h$ channels 
$c_a \simeq 2.2$, 
whereas for the fermion channels $c_t\simeq 1.7$ and $c_b\simeq 6$. 
One understands consequently why the $W$, $Z$ and $h$ channels, which give contributions $Y_{\theta}\propto M_{P}/m_{W,Z,h}$, dominate over the $b$ channel, which gives $Y_{\theta}\propto m_b^3 M_{P}/m_h^4$. As for the top channel, 
it is suppressed too because for $s\gtrsim 4 m_t^2$ it is more threshold suppressed than the other channels, see Eqs.~(\ref{sigmahiggsh})-(\ref{sigmahiggsf}), and for higher $s$ it gets quickly suppressed by the $m_t^2/s^2$ asymptotic behaviour of the top cross section (as compared to the $1/s$ asymptotic behaviour of the $W$, $Z$ and $h$ cross sections).

Let us now move to the region $m_\theta < 3$ MeV, where
the Higgs portal alone would lead to a too small relic density. In this case the $N$ annihilation process can do the job, as long as the reheating temperature is large enough to produce
a thermal population of sterile neutrinos, which we assume.\footnote{It is well-known that, if a sterile neutrino has Yukawa couplings 
inducing a light neutrino mass of the order of the atmospheric or solar neutrino mass scales, the decays/inverse decays of the sterile neutrino 
are in thermal equilibrium at $T\sim m_N$.
There are various cosmological issues which can put constraints on thermal sterile neutrinos, see e.g. \cite{Smirnov:2006bu}.
However they become relevant for sterile neutrino masses below the $\sim$~GeV scale, which are not pertinent in our case, see 
Figs.~\ref{fig2b},\ref{fig1b} below.}
The annihilation rate depends on two free parameters only, $g$ and $m_\theta$,
with the sterile neutrino mass $m_N$ determined by \eq{lagi}.
From  the freeze-in of the reaction $N N \rightarrow \theta \theta$ 
one obtains as usual $Y_{\theta}\simeq [\gamma_{annih}  /(H s)]_{T=m_N}$. 
One then finds that the DM relic density  scales as $g^6 m_\nu/m_\theta$, as confirmed by the
numerical integration of the Boltzmann equation.

Our numerical result in the $(m_\theta - g)$ plane
is shown in Fig.~\ref{fig1} for $m_\nu=0.05$~eV.
The diagonal branch of the black thick line corresponds to a freeze-in through the sterile neutrino portal,
with the correct value of the DM relic density. The required value of the $\theta-N$ coupling is given
approximatively by  $g\simeq 2 \cdot 10^{-3} (m_\theta/{\rm MeV})^{1/6} ({\rm eV}/m_\nu)^{1/6}$.
Above (below) the line the relic density is too large (small).
Since the relic density is proportional to the sixth power of $g$, the predicted value of $g$ depends very weakly 
on the cutoff $\Lambda$, which we fixed in all the numerical analysis,  $k= \log(\Lambda^2/m_N^2)/(8\pi^2) =1$.

\subsection{Non-thermal production}

In general, one may think of non-thermal, more model dependent, $\theta$ production mechanisms.
A class of mechanisms is related to the phase transition at the scale of SSB  $f$. 
At this epoch (part of) the energy stored in the scalar potential false vacuum may be transferred to the pNGBs.
There could also be cosmic strings produced at the phase transition, which would decay into pNGBs, 
but their exact density depends on model details. 
Here we assume that these contributions to $\Omega_\theta$ are negligible. This is the case, in particular,
if the universe underwent an inflationary phase at temperatures smaller than $f$ (note that the reheating temperature
can still be larger than $m_N=gf/\sqrt{2}$, since the coupling $g$ is much smaller than one in our scenario).

A potential 
 source of non-thermal production of pNGB dark matter could be provided by the oscillations of the field $\theta$ around the minimum
of its effective potential, in case the value of $\theta$ at high temperature is displaced from such minimum, a mechanism well-studied in the case of
the axion. 
Let us summarize the axion case. The
axion mass is very suppressed above $\Lambda_{QCD}$, therefore at high temperatures the axion behaves
as an exact NGB. In this case, at the end of the associated Peccei-Quinn phase transition at scale $f_{PQ}$, 
the value of the axion field $a$ can lie in any of the
equivalent vacua described by $0\le a/f_{PQ} <2\pi$. Later, when $m_a(T)$ becomes important with respect to the Hubble parameter $H(T)$,
the field $a$ begins to oscillate around zero, producing a coherent state of particles at rest with an associated $\Omega_a^{oscill}$
(for a review see Ref.~\cite{sik}).
One may naively think that this picture applies also to our scenario, 
with a negligible $m_\theta(T)$ at high temperature (where the EW symmetry is restored), and the oscillations beginning
only at $T\sim$ TeV, when $m_\theta$ is generated.
If this were the case, taking the values of $f$ and $m_\theta$ in the range we considered for the freeze-in production,
one would conclude that $\Omega_\theta^{oscill}$ overcloses the universe, 
unless the initial value of $\theta$ is tuned to be much smaller than $f$, or inflation takes place at temperatures below the TeV scale.

However, the temperature dependence of $m_\theta(T)$ is very different from the one of the axion, whose mass is generated non-perturbatevely
by an anomaly at $\Lambda_{QCD}$.
In our case,
$m_\theta$ is generated, instead, by an explicit breaking of $U(1)_X$ and therefore it does not
vanish at high temperatures. On the contrary, we expect $m_\theta(T)$ to receive large thermal corrections. 
Even if the EW symmetry is restored for $T \gtrsim$ TeV, there are contributions to $m_\theta$ that are not proportional
to the Higgs vev $v$, such as $\delta m_\theta^2 \sim g^2y^2 \Lambda^2/(16\pi^2)^2$
(see the discussion at the end of section \ref{appro}). 
These lead to $m_\theta^2(T) \simeq constant\cdot T^2$ at high
temperature. A quantitative estimate of a possible non-thermal production of $\theta$ 
would then require a non-trivial study of the thermal evolution of the field $\theta$ during and after the phase
transition at scale $f$.
Still, we just notice that $m_\theta(T)$ is typically larger than $H(T)$ already at $T\sim f$.
Thus, we argue that the field $\theta$ does not acquire a random initial value of order $f$, but rather it
sits in the minimum of the potential already at high temperature, or in other words coherent oscillations are strongly suppressed.
On the basis of this argument, we assume that $\Omega_\theta^{oscill}$ is negligible, and that the thermal freeze-in dominates
the DM relic density.

\subsection{Summary}

Our results on the $\theta$ relic density are summarized
in Fig.~\ref{fig1}, where we show the line $\Omega_\theta = \Omega_{DM}$ in the $(m_\theta - g)$ plane.
The behaviour of the line with the correct relic density can be easily understood.
For $m_\theta\simeq  0.15$~keV, $g$ must be larger than  (or equal to) the value needed to thermalize $\theta$. 
For progressively
larger $m_\theta$, the value of $g$ should correspond to less and less thermalization.
For $ 0.15$ keV $< m_\theta < 3$ MeV, one can see that the required coupling $g$ is
progressively smaller than the thermalization value, indicated by the thin diagonal line. 
When one approaches $m_{\theta}\simeq 3$ MeV, 
$g$ becomes rapidly smaller as the Higgs portal begins to contribute to the $\theta$ number density.
Once the Higgs portal produces by itself the correct amount of $\theta$-particles,
$g$ must be small enough to make the sterile neutrino freeze-in negligible.

We conclude that the observed DM relic density can be obtained thermally for 
\begin{equation}
0.15 \ \hbox{keV} \lesssim  m_\theta  \lesssim 3 \ \hbox{MeV} ~,
\end{equation}
where the exact upper value  depends on the Higgs mass, as discussed in section \ref{FI}.
Note that this upper bound  holds as an absolute prediction as soon as the Higgs portal dominates the DM production.
This is necessarily the case, in particular, if the reheating temperature lies below the sterile neutrino mass scale.
More generally,
the prediction $m_\theta\simeq 3$~MeV 
holds for any pNGB 
whose Higgs portal interaction $\lambda$ gives the dominant contribution to its mass and relic density,
independently of the associated global
symmetry and of the source of explicit symmetry breaking
which
induces $\lambda$.
In particular the SSB scale may be different from the seesaw scale.

Of course the prediction for $m_\theta$ would change
if the number of DM particles was not negligible already at temperatures higher than the EW scale.
In this case one has to produce less of them through freeze-in, which means that 
$m_{\theta}$ has to be smaller. In other words, $m_{\theta}\simeq 3$ MeV constitutes an absolute upper bound for the freeze-in mechanism, and it holds as soon as the freeze-in production dominates over the initial population.

\section{Approximate symmetries of the seesaw sector}
\label{appro}

In this section we consider the SM augmented with sterile  neutrinos $\nu^c_i$, with a global symmetry broken spontaneously at the seesaw scale,
and we discuss in some detail the generation of a mass for the associated pNGB.

Let us consider the most general Yukawa interactions to be added to the SM in the presence of gauge singlet fermions,
\beq
-{\cal L}_{\nu^c} = l_\alpha  m_{\alpha j}  \nu^c_j \left(\frac{H}{v}\right) + \frac 12 \nu^c_i M_{ij} \nu^c_j + {\rm h.c.}~,
\label{ss}\eeq
where $l_\alpha$ are the lepton doublets and $H$ is the Higgs doublet, whose neutral component acquires
a vev $v=174$ GeV.
Here and in the following, the mass parameters $m_{\alpha j}$ and $M_{ij}$ are intended in a generalized sense as dynamical
scalar fields that may or may not acquire
a non-zero vev. 
Thus, the above lagrangian has a global $U(1)_L$ symmetry
with the following lepton number assignments:
\beq
L(l)= +1~,~~~ L(\nu^c)= -1~,~~~ L(H)=L(m)= 0~,~~~ L(M)= +2~,
\eeq
where we dropped flavour indices.
When $M$ acquires a vev, $U(1)_L$ is spontaneously broken and a massless NGB appears in the spectrum of the
theory, the singlet Majoron \cite{CMP}.
One can write more explicitly
\beq
M_{ij} = g_{ij} \Phi ~,~~~~~
\Phi \equiv \frac{\rho}{\sqrt{2}} e^{i\theta/f}~,
\eeq
where $\Phi$ is a complex scalar with $L=2$, the vev $\langle \rho \rangle = f$ breaks spontaneously lepton number, and $\theta$ is the Majoron. 

There are various possible sources of $U(1)_L$ explicit breaking coming from other sectors of the lagrangian.
There may be soft terms in the scalar potential involving $\Phi$ which break lepton number; their mass scale, whose size is arbitrary, determines the induced Majoron mass (see e.g. 
\cite{GMS}). Another possible source of $U(1)_L$ explicit breaking are quantum gravity effects at the Planck scale $M_P$,
which can break in general all non-gauge symmetries. 
Assuming these effects are suppressed by powers of $M_P$, they
can be used to generate a Majoron mass at the keV scale \cite{ABST,RBS}, which has been extensively studied as DM
candidate \cite{RBS,BV,Lattanzi:2007ux,BJ}.

In this paper we will consider a different source of explicit breaking of global symmetries, 
provided by the set of the Yukawa couplings \cite{HR}.
This allows to relate the size of the pNGB parameters to the fermion mass scales already present in the theory.
We will focus on symmetries of the lagrangian in Eq.~(\ref{ss}) other than $U(1)_L$, 
with each lepton carrying in general a different charge.
In this case the symmetries are respected only by some matrix elements $m_{\alpha j}$ and $M_{ij}$,
and they are explicitly broken if some other matrix elements are non-zero.
The mass and couplings of the pNGBs will be completely determined by the seesaw parameters and
by the choice of the cutoff, since they can receive cut-off dependent quantum corrections.

The pNGB mass is, in general, quadratically sensitive to the cut-off.
In order to understand the origin of quadratic divergences and the mechanism to remove them,
consider first the explicit breaking of a $U(1)$ symmetry in a theory with only one 
sterile neutrino $\nu^c$:
\beq
-{\cal L}_{\nu^c} = \frac 12 \nu^c (M_a e^{i\theta/f}+ M_b) \nu^c + {\rm h.c.} = \frac 12 (M_a+M_b)\overline{N}N 
-\frac{i M_a}{2f} \theta \overline{N}\gamma_5 N  - \frac{M_a}{4f^2} \theta^2 \overline{N} N + {\cal O}((\theta/f)^3) ~,
\label{onef}\eeq
where $M_{a,b}$ are real mass parameters. Here $M_a$ is generated by the spontaneous breaking of the $U(1)$ symmetry associated with the
NGB $\theta$, while $M_b$ breaks this symmetry explicitly. It is instructive to compute the fermion loops generating
the pNGB mass term, 
$m_\theta^2\,\theta^2/2$,
using the right-hand side of \eq{onef}: if $M_b$ is zero, the two relevant one-loop
diagrams cancel each other, as expected for an exact NGB. However, in the presence of the explicit breaking, a non-zero
quadratically divergent contribution is left, which is given by 
\beq
m_\theta^2 \sim \frac{1}{8\pi^2} \frac{M_aM_b}{f^2} \Lambda^2~.
\eeq
The effective theory below the scale $f$ may still contain a light scalar $\theta$, since $m_\theta$ can be parametrically small, but its
NGB nature is obscured by the quadratic dependence on the details of the ultraviolet completion.

In the presence of more than one fermion family one can define several family-dependent $U(1)$ symmetries.
In general, they are explicitly broken by some mass matrix entries.
However, certain $U(1)$'s will be broken only when several entries are non-zero at the same time: 
this is the key to reduce the degree of divergence of the radiative contribution to $m_\theta$, where $\theta$'s are
the associated pNGBs \cite{HR}. 
This fact can be understood in the language of the effective potential $V_{eff}$, considering the full (active + sterile) neutrino mass matrix ${\cal M}$: 
the term in $V_{eff}$ 
quadratic in the mass matrix, $\sim {\rm Tr}({\cal M}{\cal M}^\dag)\Lambda^2$, is invariant under certain $U(1)$ symmetries, that is,
it does not depend on the associated pNGBs.
Thus, the potential of these pNGBs contains at most terms quartic in the mass matrix, $\sim {\rm Tr}({\cal M}{\cal M}^\dag{\cal M}{\cal M}^\dag)
\log \Lambda^2$, which are
only logarithmically sensitive to the cutoff (for an application of this idea to eV scale sterile neutrinos, with the pNGB playing the role of dark energy,
see \cite{BHOS}).
Of course non-abelian symmetries are also possible, with several NGBs and, potentially, qualitatively different phenomena,
but this extension will not be needed for our purposes and will not be considered in this paper.

In order to identify the combination of matrix entries that induces a pNGB mass,
we rewrite \eq{ss} 
with the replacement $l_\alpha (H/v) \rightarrow \nu_\alpha$, 
in the minimal case of two sterile neutrinos, which are sufficient for realistic light neutrino masses:
\beq
-{\cal L}_{\nu^c} = \nu_\alpha ~(m_{\alpha 1} ~ m_{\alpha 2}) \left(\ba{c} \nu^c_1 \\ \nu^c_2 \ea\right) + \frac 12 (\nu^c_1 ~ \nu^c_2) 
\left(\ba{cc} M_{11} & M_{12} \\ M_{12} & M_{22} \ea\right) \left(\ba{c} \nu^c_1 \\ \nu^c_2 \ea\right) + {\rm h.c.}~.
\label{ss12}\eeq
Here a sum over $n_f=3$ active flavours is understood ($\alpha=e,\mu,\tau$).
Let us identify the matrix entries whose phases can be removed. 
There are $2n_f + 3$ mass terms, which in general have a phase.  One can absorb $n_f + 2$
of these phases by redefining the $n_f + 2$ neutrino fields present in \eq{ss12}. 
As a result,  there remain $n_f+1$ complex matrix entries in \eq{ss12} 
(either $n_f$ phases in $m$ and one in $M$, or equivalently $n_f-1$ phases in $m$ and two in $M$). 
When one sets these $n_f+1$ entries to zero, there are no physical phases left.

Now, suppose these $n_f+1$ complex entries vanish and the non-zero entries 
are generated
by a set of  scalar fields 
$\Phi_a \equiv \rho_a \times \exp(i\theta_a/f_a) / \sqrt{2}$, 
acquiring a vev $\langle\rho_a\rangle=f_a$.
Then, by an appropriate redefinition of the lepton fields,
one can remove all the phase fields $\theta_a$ from the Yukawa lagrangian of \eq{ss12}. Thus, these fields will have only derivative couplings to the leptons, resulting from the redefined lepton kinetic terms.
This means that the lagrangian has exact $U(1)$ symmetries, broken spontaneously by $\langle\Phi\rangle$'s, and $\theta_a$ are 
exact NGBs.

When one of the zero matrix entries is switched on, however, it is no longer possible to remove all the phases. This means that one 
$U(1)$ is explicitly broken, with the associated pNGB $\theta$ acquiring non-derivative couplings. 
Thus, a non-zero $m_\theta$ is generated by neutrino one-loop diagrams.
To estimate $m_\theta$, consider the set of the ($n_f+3$) non-zero matrix entries.
One can check that $\theta$ could be rotated away from \eq{ss12} if any out of four of these entries were switched off.  
This means that a $U(1)$ symmetry is recovered
when any of them is put to zero, and therefore
$m_\theta$ must be proportional to the product of the four entries. As a consequence, quadratically divergent contributions to $m_\theta$ turn out to be absent.

The $\theta$ mass can be controlled by the entries of $M$ only, by those of $m$ only, or by both.
Let us specify three models with a $U(1)_X$ symmetry, which are representative of these three generic possibilities:
\begin{itemize}

\item[(i)]
The symmetry is broken explicitly in the singlet neutrino sector.
Two independent entries of the Majorana mass matrix $M$ are allowed by $U(1)_X$, while the third is forbidden.
For example take $X(\nu_1^c)= 0$, $X(\nu_2^c)= 1$ and one scalar field $\Phi$ with $X(\Phi)= -2$.
Then $M_{11}$ is allowed and $M_{22}$ is generated by the vev of $\Phi$, while  $M_{12} = 0$.
The NGB coupling to neutrinos reads $M_{22} \exp(i\theta/f) \nu^c_2 \nu^c_2 /2$.
(The Dirac neutrino sector is not relevant here: the $m$ entries may or may not respect $U(1)_X$, in any case
the leading symmetry breaking effects are controlled by $M$.) 
When $M_{12}$ is different from zero, neutrino loops give a non-zero contribution to $m_\theta$ as long as they
contain four appropriate mass insertions:
\beq
m_\theta^2 \sim \frac{1}{8\pi^2} \frac{M_{11}M_{12}M_{22}M_{12}}{f^2} 
\log\frac{\Lambda^2}{\mu^2}~,
\eeq
where we included a loop suppression factor and
$\mu$ is the renormalization scale, which can be taken of the order of the sterile neutrino masses.
The cancellation of quadratic divergences as well as the structure of the non-vanishing contribution to $m_\theta$
are dictated by our symmetry argument, and they can be both explicitly checked by computing the loops.

\item[(ii)] The symmetry is broken explicitly within the Dirac neutrino sector only.
The entries on a given row of the Dirac mass matrix $m$ are both allowed, while in the other rows only one entry is allowed.
For example take $X(\nu_e)=1$, $X(\nu_{\mu,\tau})=3$, $X(\nu_1^c)= -1$, $X(\nu_2^c)= 1$ and a scalar field $\phi$
with $X(\phi)= -2$.
Then $m_{e1}$ is allowed, $m_{e2}$, $m_{\mu1}$ and $m_{\tau1}$ are generated by the vev of $\phi$,
and $m_{\mu2}=m_{\tau 2}=0$.
(For simplicity we suppose that the scalar fields $\phi$ and $\Phi$ that can contribute to $m$ and to $M$ belong to different sets.
The singlet neutrino sector is assumed to respect $U(1)_X$: in the present example 
$M_{12}$ is allowed, thus giving an equal mass to the two sterile neutrinos.) 
This case reproduces closely the scenario discussed by Hill and Ross \cite{HR}, which dealt with the quark Dirac mass matrix.
Once the entries explicitly breaking $U(1)_X$ are switched on in $m$, one obtains
\beq
m_\theta^2 \sim \frac{1}{8\pi^2} \frac{m_{\alpha1}m_{\alpha2}m_{\beta 2}m_{\beta1}}{f^2}
\log\frac{\Lambda^2}{\mu^2}~,~~~~\alpha\ne \beta~.\\
\eeq

\item[(iii)] The symmetry is broken explicitly only by the interplay of the Dirac and Majorana neutrino mass matrices.
One possibility is that $U(1)_X$ allows for all the entries of $M$, e.g. taking the charges
$X(\nu_1^c)=-1$, $X(\nu_2^c)=1$ and $X(\Phi)= 2$; this symmetry can allow only one between $m_{\alpha1}$ and $m_{\alpha2}$,
depending on the charge of $\nu_\alpha$.
Another possibility is that $U(1)_X$ allows for all the entries of $m$, but it forbids two entries of $M$, 
e.g. taking the charges  $X(\nu_\alpha)=X(\nu^c_1)=0$ and $X(\nu^c_2)=-X(\phi)\ne0$.
As usual the mass of  the NGB $\theta$ is generated when one zero entry is switched on. One finds
\beq
m_\theta^2 \sim \frac{1}{8\pi^2} \frac{M_{ij}M_{kl}m_{\alpha 1}m_{\alpha 2}}{f^2}
\log\frac{\Lambda^2}{\mu^2} ~,~~~~ij\ne kl~.
\label{m3}\eeq

\end{itemize}
In order to roughly quantify the energy scales under discussion,
let us indicate with $M=gf$ ($m=yv$) a generic seesaw (electroweak) scale, possibly suppressed by a small Yukawa coupling
$g$ ($y$) with respect to the $U(1)_X$ SSB scale $f$ (the EW scale $v$), and let us
introduce also the neutrino mass scale $m_\nu =m^2/M$.
Up to a loop suppression, in the three cases 
$m_\theta$ is of order (i) $(M/f)M = g M$, at or below the seesaw scale; 
 (ii) $(m/f) m = g m_\nu$, at or below the neutrino mass scale; (iii) $(M/f)m = g m$, at or below the electroweak scale.
In the following we will concentrate on the last possibility, since we are interested in scalar DM candidates below the electroweak scale.

Before moving to the other properties of the pNGB relevant for DM, some comments are in order to
assess the soundness of the above estimate for the pNGB mass $m_\theta$.
We have seen that, in general, in the presence of explicit breaking of the associated global symmetry,
$m_\theta$ can be sensitive quadratically to the cutoff of the theory. However, if the global symmetry is 
broken only by the contemporary presence of several couplings,
then $m_\theta$ shall be proportional to the product of all these couplings.
In this case the Feynman diagrams contributing to the pNGB mass will have a lower degree of divergence
and thus quadratic divergences vanish. This possibility was often employed in model-building in the past,
and recently was extensively used in the context of little Higgs models, under the name of ``collective breaking"
(see Ref.~\cite{LH} for reviews).
The motivation is to stabilize the electroweak scale against large quantum corrections. 
In these models the Higgs is a NGB of a symmetry broken spontaneously at the scale $f_{ew}\sim$ TeV, whose mass is generated
at one-loop level by two couplings $y_{1,2}$, whose contemporary presence  breaks explicitly the symmetry.
As a consequence, $m_h^2 \sim y_1^2y_2^2 f_{ew}^2 \log(\Lambda^2/\mu^2)/(16\pi^2)$, that is, the sensitivity to the cut-off $\Lambda$
is only logarithmic. However, in general, the quadratic divergence reappears at the two-loop level, with a correction to the
Higgs mass $\delta m_h^2 \sim y_1^2y_2^2 \Lambda^2/(16\pi^2)^2$, which indicates that $\Lambda$ cannot be larger than $\sim 4\pi f_{ew}$,
in order for the theory to remain natural.

It is instructive to compare this little Higgs scenario with our scenario, where the DM candidate $\theta$ is a pNGB associated 
with SSB at the scale $f$, of the order of the
seesaw scale. In this case the pNGB mass generated at one-loop
can be written schematically as $m_\theta^2 \sim g^2 y^2 
v^2 \log(\Lambda^2/\mu^2)/(16\pi^2)$. 
The mass $m_\theta$ is not proportional to 
$f$, but rather to the electroweak scale $v$. 
Besides, it can be much smaller than $v$ because of the four powers of Yukawa couplings and the loop suppression.
This is true even with a huge cutoff $\Lambda$, since the dependence on it is only logarithmic.
However, at higher orders the quadratic divergence may reappear, e.g. through a two-loop diagram with a virtual Higgs exchanged
across the neutrino loop, leading to 
$m_\theta^2 \sim g^2 y^2 \Lambda^2/(16\pi^2)^2$.
This result is not surprising, since the mechanism we adopted explains why the
pNGB mass is related to the EW scale, but it 
does not address the stability of the EW scale against radiative corrections.
In other words, as already remarked by Hill and Ross \cite{HR}, the Higgs sensitivity to quadratic corrections 
also enters in $m_\theta$ at higher order.
In order for this two-loop correction to be negligible with respect to the one-loop estimate, one needs the cutoff of the Higgs boson loops, $\Lambda_H$, to be smaller than $\sim 4\pi v$.
(By enlarging the global symmetry, it may be possible to remove also the two-loop quadratic divergence, see e.g. \cite{BHOS}, 
but in general higher
orders reintroduce the problem.)
If one wants to stabilize a theory which includes a scale much larger than $v$ (e.g. the scale $f$),
one must address the usual hierarchy problem, e.g. postulating supersymmetry broken at $\Lambda_{susy}\sim$ TeV,
or a strongly interacting sector that condenses at $\Lambda_{c}\sim$ TeV generating dynamically the EW scale. 
In this paper we assume the stability of the EW scale is realized, and thus we can adopt our one-loop estimates 
for the pNGB mass and couplings, in order to study phenomenology.

\subsection{A seesaw model leading to the Higgs portal}

Let us consider in some detail  a specific  $U(1)_X$ symmetry of the neutrino sector, such that the pNGB $\theta$ acquires
radiatively a coupling to the SM Higgs.
Take the sterile neutrino charges $X(\nu_1^c)=-1$ and $X(\nu_2^c)=1$,
and a scalar field $\Phi$ with charge $X(\Phi)=2$, whose vev breaks the symmetry spontaneously.
The interaction lagrangian involving the sterile neutrinos and the pNGB $\theta$ reads
\beq
-{\cal L}_{\nu^c-\theta} = l_\alpha  (m_{\alpha 1} ~ m_{\alpha 2}) \frac{H}{v}\left(\ba{c} \nu^c_1 \\ \nu^c_2 \ea\right) + \frac 12 (\nu^c_1 ~ \nu^c_2) 
\left(\ba{cc} M_{11}e^{i\theta/f} & M_{12} \\ M_{12} & M_{22}e^{-i\theta/f} \ea\right) \left(\ba{c} \nu^c_1 \\ \nu^c_2 \ea\right) + {\rm h.c.}~. 
\label{ss12model}\eeq
Here $M_{12}$ is a mass term allowed by $U(1)_X$, while
$M_{11}$ and $M_{22}$ are generated after SSB. 
The symmetry is broken explicitly by either $m_{\alpha1}$ or $m_{\alpha2}$, depending on the $U(1)_X$ charge
assigned to the lepton doublet $l_\alpha$.\footnote{
The choice of $X$-charges is made to realized a model of type (iii), in accordance with the classification of the previous section.
This choice 
is not unique. An equivalent possibility is a $U(1)_X$ symmetry with charges $X(\nu_1^c)=-1$, $X(\nu_2^c)=1$, $X(l_\alpha)=0$,
broken spontaneously by a scalar field of charge $X(\phi)=1$. Then in Eq.~(\ref{ss12model}) one should replace
$m_{\alpha1} \rightarrow m_{\alpha1} e^{i\theta/f}$ and $m_{\alpha2}\rightarrow m_{\alpha2} e^{-i\theta/f}$,
while the entries $M_{11}$ and $M_{22}$ are independent from $\theta$ and represent the source of explicit symmetry breaking.
It is easy to check that the two realizations are equivalent, since the lagrangian is the same 
up to a rephasing of the neutrino fields.} 
Note that $\theta$ can be rotated away by rephasing the neutrino fields if $m_{\alpha1}=0$ or $m_{\alpha2}=0$. In this case the symmetry
is restored and $\theta$ is a true NGB. The dependence on $\theta$ can be removed also when two independent entries of $M$ are vanishing,
therefore the symmetry breaking effects must vanish also in this limit. As we discussed in the previous section, this need of four different couplings
to break the symmetry is the key to cancel quadratic divergences.\footnote{Note that the field $\theta$ can be rotated away from
\eq{ss12model} by phase redefinitions, as long as the $n_f$ entries that break explicitly $U(1)_X$ are set to zero.
This is slightly different from the general case discussed below Eq.~(\ref{ss12}), where $n_f+1$ entries must vanish, for the trivial reason
that here the same phase field appears (with opposite sign) in two independent entries.
The only physically relevant fact still holds: the product of four independent entries is needed to generate a mass for $\theta$.}

The interactions in \eq{ss12model} generate an effective potential for the pNGB $\theta$.
We assumed that $M_{ij}$ and $m_{\alpha i}$ are real and positive (see discussion below) and we
performed the one-loop computation of the effective potential, which can be written as
\beq
{\cal }V_{eff} = \frac{\lambda}{2} \theta\theta H^\dagger H +{\cal O}(\theta^4) ~. \label{leff}\eeq
We find that the quadratically divergent contributions cancel explicitly, as expected,  and the logarithmically divergent ones give
\beq
\lambda \simeq \frac{1}{4\pi^2} \frac{M_{12}(M_{11}+M_{22})}{f^2} \frac{\sum_\alpha m_{\alpha1}m_{\alpha2}}{v^2} \log\frac{\Lambda^2}{\mu^2} ~,
\label{l2}\eeq
with a renormalization scale $\mu\sim M_{ij}$ and up to tiny corrections of higher order in $m_{\alpha i}/M_{kl}$. 
This type of models is particularly predictive, because the pNGB mass is generated by the same loops that generate $\lambda$,
that is, $m_\theta$ is obtained from \eq{leff} by replacing the Higgs with its vev, $m_\theta^2=\lambda v^2$.
By taking a common value $m_N=gf/\sqrt{2}$ ($m=yv$) for each entry of the Majorana (Dirac) neutrino mass matrix, \eq{l2} reduces to \eq{mesti},
up to a factor $2n_f$ accounting for the sum over flavour indexes.
As we saw in section \ref{relic}, the value of the coupling $\lambda$ has a crucial role for the (partial) thermalization of $\theta$ and thus for the determination of its relic density.

A comment  on the CP symmetry is in order. In general, the mass terms $M_{ij}$ and $m_{\alpha i}$
may be complex, that is, they may carry phases that cannot be removed by a redefinition of the fields. These phases correspond to an explicit
breaking of CP. In this case $\theta$ would have both scalar and pseudo-scalar couplings to the fermions, with characteristic
phenomenological signatures \cite{HR}. 
Moreover, the effective potential would contain terms odd in $\theta$, such as $\mu\theta H^\dag H$ (with $\mu \ll v$), 
that induce a vev for $\theta$ and a small
mixing with the Higgs. This may endanger the stability of $\theta$, since it would couple linearly to SM particles. 
Still, these couplings can be very small
and in addition the $\theta$ decays may be kinematically forbidden if $m_\theta$ is sufficiently small.
In this paper we do not investigate this more complicated CP-violating possibility,
and we rather assume that CP is a good symmetry of the seesaw sector.
In this case  $M_{ij}$ and $m_{\alpha i}$ 
in \eq{ss12model}
are all real and 
$\theta$  preserves its pseudo-scalar nature, since  \eq{ss12model}
is invariant under a CP transformation with $\theta \rightarrow -\theta$. 
Usually the DM stability requires an additional (discrete) symmetry that forbids its decay into SM particles.
In the present scenario the CP symmetry itself guarantees that $\theta$ couples linearly only to the heavy sterile neutrinos, and we will see in section \ref{intau} that this makes $\theta$ sufficiently long-lived.

\section{Constraints on the pNGB lifetime \label{intau}}

We first derive the couplings of $\theta$ to the SM fermions and gauge bosons, and compare its decay width into light SM particles with the
lifetime of the universe. 
Subsequently, assuming that $\theta$-particles account for the whole DM density,
we discuss the more stringent astrophysical and cosmological constraints which exist on the decay of $\theta$ to 
neutrinos, electrons and photons.

\subsection{$\theta$ couplings to the SM particles}

The pNGB lifetime is determined by its couplings to light fermions. These come from the $\theta$-$N$ interaction, through 
the $\nu$-$N$ mixing induced by the Dirac neutrino masses. 
In this way $\theta$ can decay to light neutrinos (at tree level) and to charged fermions (at one-loop level).
In turn, these couplings to SM fermions could induce, through triangle loop diagrams, couplings to SM gauge bosons. 
These decays should be sufficiently slow to make the DM lifetime longer than the age of the universe,
$\tau_0 \simeq 5\cdot 10^{17}$s.
This provides interesting constraints on the seesaw parameters, as we now discuss.

\begin{figure}[t]
\begin{center}
\includegraphics[width=13cm]{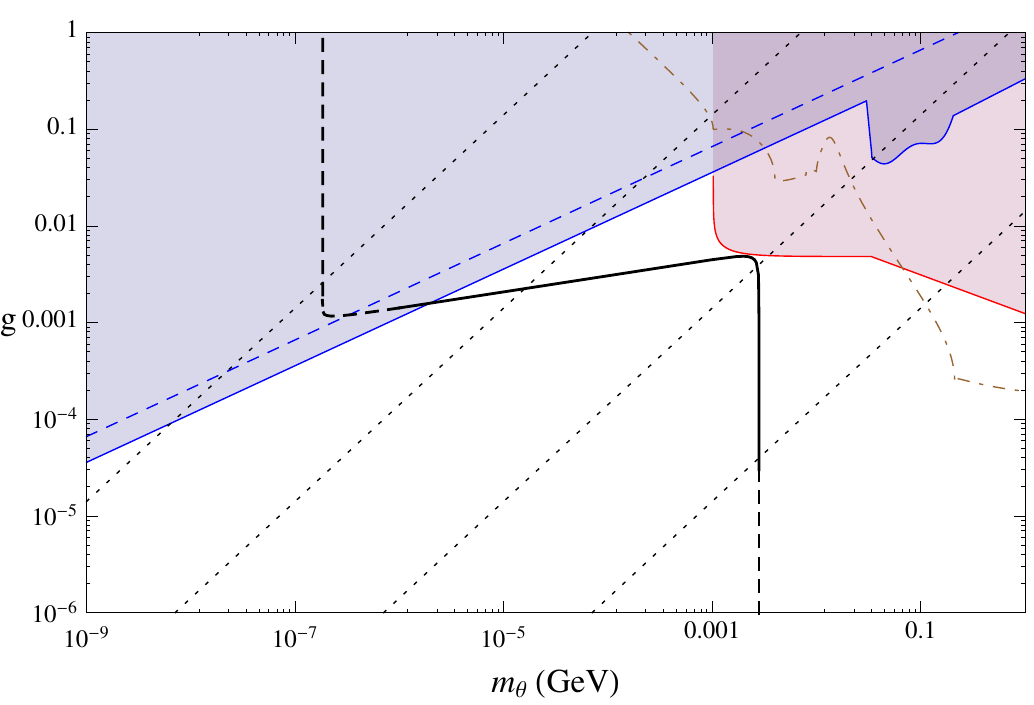}
\caption{\small The constraints on the DM lifetime in the $(m_\theta - g)$ plane, for $m_\nu=0.05$ eV. The thick black curve, corresponding to
the correct DM relic density, as well as the dotted lines, corresponding to constant values of $m_N$,
were already presented in Fig.~\ref{fig1}. The curve is dashed for $m_\theta \lesssim 1$ keV, because
DM is warm in this region (see the text), and for $g\lesssim 10^{-5}$,
because of the theoretical bound $f<M_P$, shown in Fig.~\ref{fig1}. 
The blue solid (dashed) line is the upper bound on $g$ from 
DM decays into neutrinos coming from astrophysics and cosmology
(from the universe lifetime): the blue shaded region is therefore excluded. 
The red solid line is the analog bound for DM decays into $e^+e^-$: the red shaded region is correspondingly excluded.
Finally, the brown dot-dashed line is a conservative estimate of the upper bound on $g$ from DM decays into photons (see the text).}
\label{fig2b}
\end{center}
\end{figure}

\begin{figure}[t]
\begin{center}
\includegraphics[width=13cm]{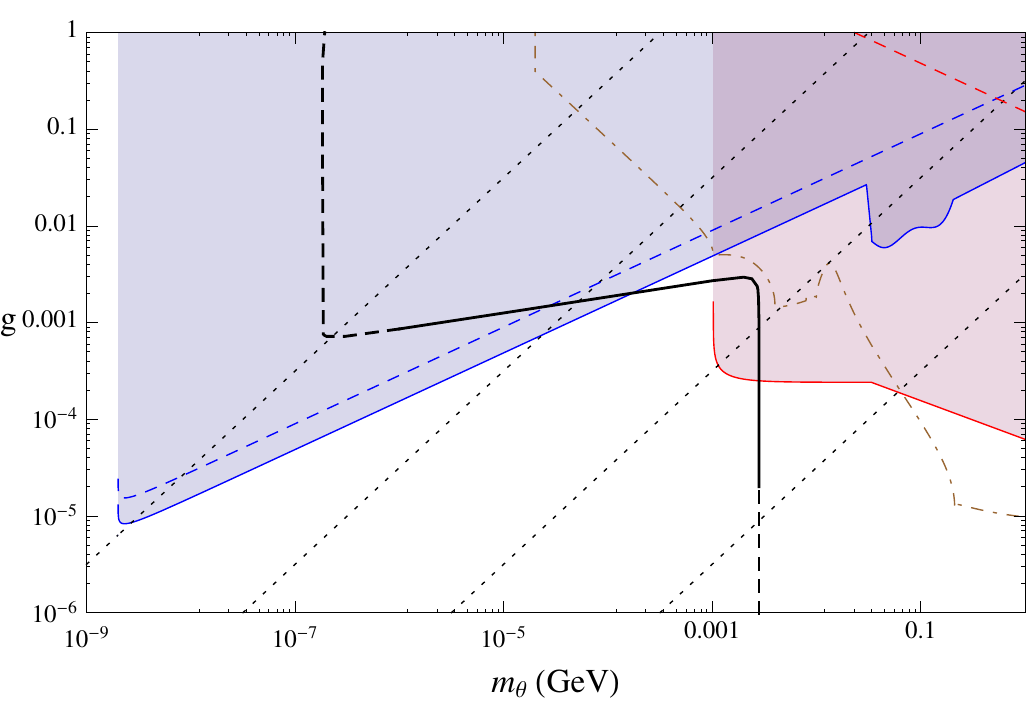}
\caption{\small The same as in Fig.~\ref{fig2b}, but for a heavier neutrino mass scale, $m_\nu=1$ eV.
The extra red dashed line is the upper bound from $1/\Gamma(\theta\rightarrow e^+e^-)>\tau_0$ (in the case of Fig.~\ref{fig2b}
the analog line lies entirely in the region $g>1$ and $m_\theta>1$ GeV, which is not displayed).}
\label{fig1b}
\end{center}
\end{figure}

{\bf Neutrinos.} At tree level, $\theta$ couples only to light neutrinos as follows:
\beq
{\cal L}_{\theta\nu\nu} = \frac{i}{2} \sum_{\alpha\beta} \frac{(\mu_\nu)_{\alpha\beta}}{f} \overline{\nu_\alpha}\gamma_5 \nu_\beta \theta ~,~~~~
(\mu_\nu)_{\alpha\beta} \equiv -\sum_{ij}  X_{ij} (mM^{-1})_{\alpha i}M_{ij} (M^{-1}m^T)_{j\beta} ~.
\eeq
Here $X_{ij}$ is the power of $e^{i\theta/f}$ associated with the sterile neutrino mass matrix entry $M_{ij}$. In the singlet Majoron model
$X_{ij}=1$ for all $i,j$, therefore one finds $\mu_\nu = m_\nu \equiv -m M^{-1} m^T$, which is the usual seesaw formula. 
On the contrary, in our models based on a family-dependent $U(1)_X$ symmetry, $\theta$ couples differently to each entry of $M$,
as for example in the model of \eq{ss12model}.
This lagrangian
leads to a total decay width (into both neutrinos and antineutrinos) 
\beq
\Gamma(\theta\rightarrow \nu\nu)=\frac{1}{16\pi}g^2_{\theta\nu\nu} m_\theta~,~~~~~g_{\theta\nu\nu}^2 \equiv 
\frac{{\rm Tr}(\mu_\nu^\dag \mu_\nu)}{f^2}~,
\label{width_nunu}
\eeq
where we neglected $(m_{\nu i}/m_\theta)^2$ corrections in the phase space, $m_{\nu i}$ being the light neutrino mass eigenvalues. 
In the Majoron case one obtains $g_{\theta\nu\nu}^2 = \sum_i m_{\nu i}^2 / f^2$.
This width is smaller than $1/\tau_0$ 
for 
$g_{\theta\nu\nu}\lesssim 3 \times10^{-19} ({\rm MeV}/m_\theta)^{1/2}$.
Such tiny coupling is natural for $\theta$,
because the pNGB couplings are suppressed by the SSB scale $f$: one needs 
$f \gtrsim 3 \times 10^9$GeV$(m_{\nu}/$eV$)(m_\theta/$MeV$)^{1/2}$,
where we used the one-family approximation,   $g_{\theta\nu\nu}\simeq m_{\nu}/f$.
To translate this bound in the $(m_\theta-g)$ plane we insert the relation $f=\sqrt{2}m_N/g$ in Eq.~(\ref{lagi}), which gives
\begin{equation}
\label{thetanunu_plot}
g_{\theta\nu\nu} \simeq 10^{-21}  \left( \frac{\rm MeV }{m_\theta}  \right)^{2} \left(\frac{g}{10^{-3}}\right)^3 
\left( \frac{m_\nu }{{\rm eV}}  \right)^{2} k~.
\end{equation}
The condition $1/\Gamma(\theta\rightarrow \nu\nu) > \tau_0$ excludes the region above the blue dashed line in Figs.~\ref{fig2b},\ref{fig1b}.

{\bf Charged fermions.}
The pNGB $\theta$ couples also to charged fermions,  through EW one-loop diagrams.
This effect arises because of the mixing between the sterile and the weakly-interacting neutrinos,
in particular it is also operative in the singlet Majoron model \cite{CMP,Pilaftsis}.
The coupling to quarks is induced by a one-loop $\theta-Z$ mixing diagram, with neutrinos in the loop.
The coupling to charged leptons is generated by an analog $Z$-exchange diagram plus a triangle diagram with $W$-exchange.
The resulting coupling can be written in a compact form in
the one-family approximation, as follows:
\beq
{\cal L}_{\theta f \bar{f}} = i g_{\theta f \bar{f}} \overline{f}\gamma_5 f \, \theta ~,~~~~
g_{\theta f\bar{f}} = \pm\frac{G_F}{(4\pi)^2} \frac{\sqrt{2}m_N}{f} m_f m_\nu ~,
\label{Ltff}\eeq
where the sign is $+$ ($-$) for up quarks and charged leptons (down quarks), and $G_F$ is the Fermi coupling constant.
This effective coupling can be suppressed only taking a small $g= \sqrt{2}m_N/f$, since all the other parameters in 
$g_{\theta f\bar{f}}$ can be determined experimentally. 
The decay width is given by
\beq
\Gamma(\theta\rightarrow f \bar{f})=\frac{1}{8\pi}g^2_{\theta f \bar{f}} m_\theta \left(1-\frac{4m_f^2}{m_\theta^2}\right)^{1/2}~.
\label{width_ff}
\eeq
We remark that all models
where the singlet Majoron is given a mass larger than one MeV and plays the role of DM candidate are constrained (or 
already excluded) by such decays into charged fermions.

In our scenario, the requirement $1/\Gamma(\theta\rightarrow f\bar{f}) > \tau_0$ leads to an upper bound
 $g\lesssim 2 ({\rm MeV}/m_\theta)^{1/2}({\rm eV}/m_\nu)$ $({\rm MeV}/m_f)$, where we have assumed $m_\theta \gg 2m_f$ for simplicity.
This bound excludes the region above the red dashed line in Fig.~\ref{fig1b}. 
As a consequence, taking into account that 
$g$ and $m_\theta$ are related by \eq{mesti}, and that  
the Dirac mass $m=yv$ cannot be larger than $\sim$ TeV, we conclude that $m_\theta$ close to the EW scale would imply 
a lifetime shorter than $\tau_0$. 
In fact, even stronger constraints on the decay width into 
charged fermions 
come from astrophysical and cosmological observations (see section \ref{CAC}), which will lead to a stronger constraint, $m_\theta < 1$ GeV.
In particular, in this class of models $\theta$ cannot be the $\sim 50$ GeV DM candidate produced by the freeze-out
of the $\lambda$ interaction with the Higgs, discussed in section \ref{fo}.
One is left with the possibility of a sub-GeV DM candidate, because in this case the decays into charged fermion pairs
are sufficiently slow (or forbidden kinematically), and the correct relic density can be generated by the freeze-in mechanism.

In the realistic three-family case, the coupling $g_{\theta f\bar{f}}$ is generically of the same order, but with a complicated
dependence on flavour parameters.
In particular one may argue that some cancellation can take place, to reduce the $\theta$ decay width.
In addition, we have seen in section \ref{appro} that the pNGB mass generation depends crucially 
on the interplay between the flavour structures of the matrices $m$ and $M$. 
Therefore, if one were to study the whole parameter space, one should 
know the explicit dependence of $g_{\theta f\bar{f}}$ on the mass matrix entries.

For illustration, we display the result for the model in \eq{ss12model}, 
considering for simplicity only one lepton doublet. Writing the Dirac mass matrix as $m\equiv (m_1~m_2)$
and keeping terms up to order $(m_i/M_{jk})^2$,
the effective coupling of $\theta$ to fermions through the mixing with the $Z$ gauge boson is given by 
\begin{equation}
{\cal L}^Z_{\theta f\bar{f}} = i \left( \frac{M_{11}}{f} F_1 - \frac{M_{22}}{f} F_2\right) 
\frac{2 \sqrt{2}\ G_F}{(4 \pi)^2} \ m_f \ T_{3f}  \bar f \gamma_5 f  \,  \theta~,
\label{tz2}\end{equation}
\begin{equation}
F_1 = m_{   1}^2 \left( c^2 s^2 K + \frac{c^2}{M_1} +  \frac{s^2}{M_2} \right) 
- m_{    2}^2\  c^2 s^2 K
- m_{    1} m_{    2}\  c s \left( (c^2-s^2) K - \frac{1}{M_1} + \frac{1}{M_2} \right) ,\,
\end{equation}
\begin{equation}
F_2 =  m_{    2}^2 \left( c^2 s^2 K + \frac{s^2}{M_1} +  \frac{c^2}{M_2} \right) 
- m_{    1}^2\  c^2 s^2 K 
+ m_{    1} m_{    2} \ c s \left( (c^2-s^2) K + \frac{1}{M_1} - \frac{1}{M_2} \right) ,\,
\end{equation}
where $T_{3f}$ is the third isospin component of the left-handed part
of the fermion $f$, $M_{1,2}$ are the eigenvalues of the matrix $M$,
which is diagonalized by a rotation of angle $\delta$, defined by $\tan 2\delta =2M_{12}/(M_{11}-M_{22})$, and we
denoted $c\equiv\cos\delta$ and $s\equiv\sin\delta$.
Finally, the loop function $K=K(M_1,M_2)$ is given by 
\beq
K(M_1,M_2)\equiv -\dfrac{M_1^2+4M_1M_2+M_2^2}{M_1M_2(M_1+M_2)}
+\dfrac{4(M_1^2+M_1M_2+M_2^2)}{(M_1-M_2)(M_1+M_2)^2} \log\dfrac{M_1}{M_2} ~.
\label{K}\eeq
Concerning the $\theta$ decay into charged leptons, one needs to add the contribution of the $W$-exchange diagram.
The main phenomenological constraint comes from 
$\theta \rightarrow e^+ e^-$. Therefore, in our simplified calculation we identify the
active neutrino with the electron neutrino $\nu_e$. 
Then, the additional  contribution to the effective coupling of $\theta$ to electrons is given by
\begin{equation}
{\cal L}^W_{\theta e\bar{e}} = i \left( \frac{M_{11}}{f} F_1 - \frac{M_{22}}{f} F_2\right) 
\frac{2 \sqrt{2}\ G_F}{(4 \pi)^2} \  m_e   \bar e \gamma_5 e \,  \theta~,
\label{tw2}\end{equation}
that carries a relative factor $-2$ with respect to the $Z$-exchange contribution. 
This is why the sum of the two contributions in Eqs.~(\ref{tz2}) and (\ref{tw2}) gives $g_{\theta e \bar{e}}/m_e = g_{\theta u \bar{u}}/m_u =
- g_{\theta d \bar{d}}/m_d$, consistently with the one-family result in \eq{Ltff}.
Also, the functions $F_{1,2}$ are of the order $m_i^2/M_j \sim m_\nu$, so the couplings are of the same order as in \eq{Ltff}.
Still, cancellations between the various terms
are possible leading to a suppression of the $\theta$ decay width for special flavour structures.
This possibility may deserve a future investigation, since an appropriate family symmetry
could in principle raise the pNGB lifetime, so that $\theta$ could become a viable DM candidate
even for $m_\theta > 1$ GeV, a region where freeze-out could lead to the observed relic density.
In this paper we do not invoke a family symmetry for the suppression of $g_{\theta f \bar f}$ to happen, 
but rather we rely on the one-family estimate 
given in \eq{Ltff}.

{\bf Gauge bosons.} 
The last effective coupling of the pNGB that may have important phenomenological consequences is the one to photons.
Let us discuss first the limit of exact global symmetry, and then comment on the effect of explicit breaking.
In general, the couplings of a NGB to gauge bosons are controlled by the gauge anomalies of the associated global symmetry.
In the case $\theta$ is the Majoron, the global symmetry can be identified with $B-L$, because $\theta$ can be rotated away
from the Yukawa interactions by rephasing all the SM fermions with a $B-L$ transformation.
Since $B-L$ is anomaly-free with respect to the SM gauge symmetries, this rephasing does not generate anomalous couplings
of the type $\theta F\tilde{F}$, rather the only leftover interaction is derivative, $(\theta/f) \partial_\mu J^\mu_{B-L}$, which is generated
by the redefinition of the fermion kinetic terms. One concludes that the Majoron  has no anomalous
couplings to gauge bosons.\footnote{
Equivalently, one may remove $\theta$ from the Yukawa interactions by an $L$ transformation; in this case one is left
with the couplings $(\theta/f) [\partial_\mu J^\mu_{L} + (c_W F_W \tilde{F}_W+c_Y F_Y \tilde{F}_Y)/(16\pi^2)]$, 
where $c_{W,Y}\ne 0$ account for the EW anomalies of the
lepton number symmetry. There is no physical difference between the two pictures. In particular one can check that
(i) $\theta$ does not couple
to weak gauge bosons at one-loop even in the $L$ case, because
the triangle diagrams generated by the derivative coupling to $J^\mu_L$ cancel
$c_{W,Y}$ exactly; (ii) $\theta$ does not decay to two on-shell quarks even in the $B-L$ case,
because the quarks are vector-like under $B-L$.}

This discussion is easily generalized to the $U(1)_X$ symmetry that we consider in this paper. First, note that the sterile neutrino charges are 
irrelevant for
gauge anomalies. In order to allow for fermion masses, we can take $X(l_\alpha)=-X(e^c_\alpha)=x_\alpha$ for $\alpha=e,\mu,\tau$, and zero $X$-charge
for the quarks. Then it is easy to check that $U(1)_X$ is anomaly-free with respect to electromagnetism and colour, and thus $\theta$ does not
couple to photons and gluons. The anomaly with EW interactions is proportional to $\sum_\alpha x_\alpha$ and can also be taken equal to zero for simplicity.
In summary, in the limit of exact symmetry our NGB does not couple to gauge bosons.

It is far more difficult to compute the pNGB couplings to gauge bosons in the presence of explicit symmetry breaking sources.
One expects that such couplings will arise at some level, since there is no symmetry arguments to prevent them at all orders. However,
the computation of the lowest order non-vanishing contribution is non-trivial. We have shown explicitly above that, in the present framework,
$\theta$ couples at tree-level to neutrinos only, while, at one-loop, couplings to the Higgs boson as well as to charged fermions are induced.
At two-loop there is a number of diagrams that connect $\theta$ to two gauge bosons. These diagrams do not necessarily add up to zero:
on the one hand, they involve the neutrino mass parameters that break explicitly $U(1)_X$, thus they are not expected to respect the anomaly
argument above; on the other hand, however, 
these parameters may not be sufficient to induce an operator $\theta F\tilde{F}$ already at two-loop order.

The computation of these two-loop diagrams is beyond the purpose of the present paper, because we will show that a
decay $\theta\rightarrow \gamma\gamma$ induced at this order would be irrelevant for phenomenology anyway: once the constraints from 
$\theta\rightarrow f \bar{f}$ are imposed, the surviving DM parameter space is not further reduced by constraints on photons. 
To see this, we estimate the size of a two-loop contribution by taking the
effective one-loop couplings in \eq{Ltff}, and computing the usual fermion triangle diagrams with two final state photons. 
The effective lagrangian can be written as
\beq
{\cal L}_{\theta\gamma\gamma} = \frac 18 g_{\theta\gamma\gamma} \theta \epsilon_{\mu\nu\alpha\beta} F^{\mu\nu} F^{\alpha\beta} 
\eeq
and the corresponding decay width is
\beq
\Gamma(\theta\rightarrow \gamma\gamma) = \frac{1}{64\pi} g_{\theta\gamma\gamma}^2 m_\theta^3~,~~~~~~
g_{\theta\gamma\gamma}^{estimate} \simeq \frac{\alpha}{\pi} \sum_f \frac{g_{\theta f \bar{f}}}{m_f} G\left(\frac{m_\theta}{2m_f}\right)~,
\label{width_gg}\eeq
where $G(x)=|\arcsin^2 x|/x^2$ \cite{ABBM}. Note that even in the limit of exact NGB, with $m_\theta \ll m_f$ and thus $G(x)\approx 1$, 
the sum does not give zero when the sign in \eq{Ltff}
is taken into account. Therefore, 
we stress again that this is just a conservative estimate, and the computation of all the two-loop contributions
may lead to a further cancellation.
Neglecting the mass dependence in $G$, the requirement $\Gamma(\theta\rightarrow \gamma\gamma)<1/\tau_0$ gives the order of magnitude
constraint
$g\lesssim 4 \times 10^3({\rm MeV}/m_\theta)^{3/2}({\rm eV}/m_\nu)$.
This bound is weaker than the one obtained below \eq{width_ff} from $\Gamma(\theta\rightarrow f\bar{f})$,
as long as $m_\theta \lesssim 1$ GeV, which is the region relevant in the present scenario.

\subsection{Cosmological and astrophysical bounds on $\theta$ couplings \label{CAC}}

In practice, the DM lifetime has to be larger than the lifetime of the universe, because late DM decays
affect several cosmological and astrophysical observations.
To derive the corresponding bounds on our scenario we make the assumption that $\theta$ is all the DM in the universe.
For the sake of simplicity, here too we take the approximation of one lepton family, 
barring large cancellations between the flavour parameters.
A sub-GeV particle could decay into neutrinos, electrons and photons. We discuss these decay channels in turn.

Let us first consider the decay $\theta \rightarrow \nu \nu$.  As it was noted in
Ref.~\cite{Lattanzi:2007ux} such a decay could affect the expansion history of the universe, because it represents energy transfer from a non-relativistic $\theta$ to relativistic neutrinos. Using SNIa and CMB data, one obtains the bound \cite{Gong:2008gi}
\begin{equation} 
\label{nunuexp}
\Gamma (\theta \rightarrow \nu \nu) < 4.5 \times  10^{-20} \ {\rm s}^{-1} \quad ({\rm or} \ \tau > 700 \ {\rm Gyr}) ~.
\end{equation}
Using \eq{width_nunu}, the corresponding bound on the coupling of $\theta$ to neutrinos is
\begin{equation}
g_{\theta\nu\nu} < 4 \times 10^{-20} \left( \frac{ {\rm MeV} }{m_\theta}  \right)^{1/2} ~.
\label{tnn}
\end{equation}
This upper bound 
can be improved in the range 30 MeV $< m_\theta< 200$ MeV \cite{PalomaresRuiz:2007ry}, using searches for the diffuse neutrino supernova background by Super-Kamiokande. For masses $m_\theta > 200$ MeV the best limit come from atmospheric neutrino observations. Since the observed spectrum coincides with theoretical estimates, one can set an upper bound on $g_{\theta \nu \nu}$ \cite{PalomaresRuiz:2007ry,Bell:2010fk}.
We use the relation in \eq{thetanunu_plot} to translate these constraints in the $(m_\theta-g)$ plane.
The blue region 
shown in Figs.~\ref{fig2b},\ref{fig1b} is excluded by the combination of the observational bounds discussed above.

Let us now consider the $\theta \rightarrow e^+ e^-$ decay.
We adapt to our case the analysis performed  in Ref.~\cite{Bell:2010fk}.  For $m_\theta \leqslant$ 20 MeV, the dominant physical process 
which constrains the parameter space is annihilation at rest, contributing to the 511 keV line. The limit is approximately given by
\begin{equation}
\label{annihilation_rest}
\left(   \frac{ {\rm MeV} }{ m_\theta  } \right)   \Gamma (\theta \rightarrow e^- e^+) < 5 \times  10^{-27}\ {\rm s}^{-1} ~.
\end{equation}
Using \eq{width_ff}, this limit leads to 
\begin{equation}
g_{\theta e e} <  9 \times  10^{-24}  \left( 1 - \frac{4 m_e^2}{ m_\theta^2} \right)^{-1/2} ~.
\end{equation}
For 20 MeV $\leqslant m_\theta \leqslant$ 1 GeV the dominant process is 
internal bremsstrahlung, i.e., photons radiated from the final electron or positron. 
One gets a slighly more stringent bound than the one shown in \eq{annihilation_rest} \cite{Bell:2010fk}.
The predicted value of the coupling $g_{\theta e e}$ in our model, as follows from \eq{Ltff}, is
\begin{equation}
g_{\theta e e} = g \, \frac{G_F}{(4\pi)^2} \, m_\nu m_e = 3 \times 10^{-23} \left(\frac{g}{10^{-3}}\right)  \frac{m_\nu}{\rm eV} ~.
\end{equation}
The corresponding observational bound is plotted in Figs.~\ref{fig2b},\ref{fig1b} as the red shaded region.

Note that the value $m_\theta\simeq 3$~MeV, predicted by the Higgs portal, could 
lead to an excess of 511 keV $\gamma$ rays from the galactic center of the Milky Way.  
As discussed in Ref.~\cite{Prantzos:2010wi}, 
a $\sim 1.5$~MeV electron or positron, produced by a $\theta \rightarrow e^+ e^-$ decay,
is enough non-relativistic to subsequently annihilate at rest in the galactic center, leading to a $\gamma$ line at 511 keV.
In particular, with such a mass one can easily obtain a $\gamma$ flux excess of the order of the one observed by the INTEGRAL $\gamma$ 
observatory \cite{Prantzos:2010wi,Weidenspointner:2008zz}.
However, unless the DM galactic profile is much more cuspy than the usually considered profiles, a DM decay gives a flux 
that is not sufficiently peaked around the galactic center \cite{Prantzos:2010wi,Ascasibar:2005rw} to be able 
to reproduce the morphology of the signal observed by INTEGRAL.

We discuss now the decay $\theta \rightarrow \gamma \gamma$.
For small masses, $m_\theta \lesssim$ keV, the photons energy is absorbed by the baryonic gas in the early universe and the processes of 
recombination and reionization are affected.  One can use the analysis done in Ref.~\cite{Zhang:2007zzh} 
(see also \cite{Chen:2003gz}) to bound the coupling  $g_{\theta\gamma \gamma}$.
For larger masses, $m_\theta \gtrsim $ keV, photons are no longer absorbed and propagate freely. Their contribution to the isotropic diffuse photon background allows also to bound  $g_{\theta\gamma \gamma}$ \cite{Chen:2003gz,Yuksel:2007dr}. 

For the region we are most interested in, stronger bounds can be obtained 
from the gamma-ray line emission limits from the Milky-Way central region.
Indeed, in the range 40 keV$  < m_\theta  <$16 MeV
\cite{Yuksel:2007dr}, one obtains
\begin{equation} 
\Gamma (\theta \rightarrow \gamma \gamma) <  10^{-28} \, {\rm s}^{-1} \, \left( \frac{ m_\theta  }{ {\rm MeV} }  \right) ~.
\end{equation}
Using \eq{width_gg}, this corresponds to
\begin{equation} 
g_{\theta\gamma \gamma} < 4 \times 10^{-21}   \left( \frac{ {\rm MeV} }{m_\theta}  \right) {\rm GeV}^{-1}  ~.
\label{tgg}
\end{equation} 
Taking the estimate for $g_{\theta\gamma \gamma}$ given in \eq{width_gg},  all these constraints translate in an upper bound on $g$,
which is shown in Figs.~\ref{fig2b},\ref{fig1b} by the brown dot-dashed curve.

We should mention that there are astrophysical constraints on the couplings 
$g_{\theta\nu\nu}$ \cite{Grifols:1988fg}, $g_{\theta e e}$ \cite{Raffelt:1996wa}, and  $g_{\theta\gamma\gamma}$  \cite{Raffelt:1996wa, Masso:1995tw}, 
based on stellar energy loss due to $\theta$ emission, provided $m_\theta$ is low enough to be produced in stellar interiors. These limits are valid without the need to assume that $\theta$ is DM. However, these astrophysical limits  are in general weaker than the ones mentioned above.

We also would like to note that the region we consider for pNGB DM masses includes the keV scale, corresponding to warm dark matter (WDM).  
Since at the epoch of structure formation WDM has free-streaming lengths below the Mpc scale, having WDM at least as a non-negligible 
DM component can alleviate some of the disagreement between the standard cold DM scenario and a variety of galactic observations 
at small scales \cite{Moore:1999nt}. The ideal observation which could place a limit on WDM is Lyman-$\alpha$ forest, i.e. 
the Lyman-$\alpha$ absorption produced by intervening neutral hydrogen in the spectra of light emitted by distant quasars. 
Using Ly-$\alpha$ observations together with other cosmological data sets, different groups 
have put lower limits on $m_{WDM} $, assuming WDM has a thermal distribution and that it is the whole of DM.
The bounds in the literature \cite{Viel:2005qj}
differ by factors of a few, ranging from 0.5 keV up to 4 keV. 
In our case, the DM species is not in thermal equilibrium and the bounds can slightly
change compared to the WDM thermal relic case.
In addition, if WDM is only a part of the whole of DM the bounds are relaxed.
We should mention that there are some potential problems in the obtention of these bounds, and they should be regarded as controversial. 
On the one hand there could be large systematic errors, and on the other hand the Ly-$\alpha$ analysis has to be performed at scales 
which are already in the non-linear regime, where calculations are less reliable. If one disposes of the Ly-$\alpha$ data these bounds 
on WDM disappear altogether \cite{Hannestad:2003ye}.

\section{Conclusions}

We proposed a new pseudo-scalar gauge-singlet DM candidate $\theta$ with mass in the keV - MeV range.  
Its couplings to the SM particles are feeble, because they are mediated by new physics at a large scale $f$, 
which we identify 
with the seesaw scale. 
The $\theta$ relic density can be produced by the scattering with the heavy particles (the sterile neutrinos), 
at temperatures of the order of $f$, or alternatively it can be generated at the EW scale through the Higgs portal,  
by the tiny $\theta$-$H$ coupling $\lambda \sim 10^{-10}$,
which is induced by the seesaw interactions.
Today, $\theta$ decays into light neutrinos and, if it is heavier than one MeV, into $e^+ e^-$, with rates that can saturate the
present upper bounds.

We argued that such a candidate is theoretically well-motivated. The heavy new physics sector is generically associated with several global 
symmetries. Some of the corresponding pNGBs may remain light, if the explicit symmetry breaking effects are sufficiently small.
For concreteness, we demonstrated that a $U(1)_X$ family symmetry of the sterile neutrino sector can be broken collectively by a set of
neutrino Yukawa couplings, so that the pNGB mass $m_\theta$ is proportional to the EW scale and to the
(small) product of the Yukawa couplings. In such scenario $m_\theta$ is not quadratically sensitive to the cutoff 
(at leading order),
therefore the presence of a light DM scalar below the EW scale is justified.

In order to calculate the $\theta$ relic density, 
we computed the rate of the $\theta$ interactions with a sterile neutrino $N$  and, through the Higgs portal, with the SM particles,
and we studied numerically the Boltzmann equation for the $\theta$ number density. 
In this framework there are only two independent parameters, 
the mass of the pNGB, $m_\theta$, and the coupling of the pNGB to the sterile neutrino, $g$, a feature 
which makes our scenario especially predictive.
We find that the Higgs portal produces the desired DM relic density, through the freeze-in mechanism, 
for a unique value of the DM mass, $m_\theta\simeq 3$ MeV. 
This prediction relies only on the relation between the DM mass and its coupling to the Higgs, $m_\theta^2 
= \lambda v^2$, which is well-justified in the case of our pNGB.
As long as the reheating temperature is sufficiently high, the sterile neutrino portal 
can produce the required relic density by freeze-in for smaller values of $m_\theta$, from $\simeq 3$ MeV
down to $0.15$ keV. 
These results are summarized in Fig.~\ref{fig1}.

The constraints 
from 
$\Gamma(\theta \rightarrow \nu\nu)$, however, exclude the region $m_\theta \lesssim 1(100)$ keV, for a light
neutrino mass $m_\nu = 0.05(1)$ eV.
In addition, the constraints
from 
$\Gamma(\theta \rightarrow e^+ e^-)$ put an upper bound on the 
coupling $g$ in the region $m_\theta \gtrsim 1$ MeV:
for $m_\nu = 0.05(1)$ eV and $g \sim 10^{-3}(10^{-4})$, the expected electron-positron flux is close to the present sensitivity.
These results are summarized in Figs.~\ref{fig2b},\ref{fig1b}.

In turn, the seesaw scale can be constrained, by requiring
our pNGB  to be a viable DM candidate in agreement with all the bounds above. 
In the case of freeze-in through the sterile neutrino portal, we find 
$10^5$ GeV $\lesssim m_N \lesssim 10^{10}$ GeV, and a SSB scale $f \sim 10^3 m_N$.
In the case of the Higgs portal, one has instead  $10^{10}$ GeV $\lesssim m_N \lesssim 10^{14}$ GeV, and 
correspondingly $10^{13}$ GeV $\lesssim f \lesssim M_P$.

Finally, we note that our analysis of the parameter space was mostly performed in the one-family approximation, that is, neglecting
possible hierarchies among the sterile neutrino mass parameters and the neutrino Dirac Yukawa couplings. 
Some of our main results are independent from this approximation, such as the allowed range for the DM mass.
On the contrary, the bounds on $g$ and $m_N$ are obviously sensitive to the flavour structure. A rough idea of these dependence
can be grasped by comparing the case $m_\nu=0.05$ eV (Fig.~\ref{fig2b}) with the case $m_\nu=1$ eV (Fig.~\ref{fig1b}).
A more detailed exploration of the flavour parameter space of this scenario will be desirable, in particular, if one wants to compare
with neutrino flavour models.

\section*{Acknowledgements}

It is a pleasure to thank K.S.Babu, X.Chu, J.R.Espinosa, D.Greynat, K.Jedamzik, S.Lavi\-gnac, 
S.Peris, S.Palomares-Ruiz, A.Pomarol, M.Quiros, J.Re\-dondo, J.Serra, M.H.G.Tytgat and A.Vara\-gnolo 
for very useful discussions. 
The work of MF was supported
in part by the Spanish Research Project CICYT-FEDER-FPA-2008-01430, the
Consolider-Ingenio 2010 Programme CPAN (CS-D2007-00042), and
the Marie-Curie Reintegration Grant PERG06-GA-2009-256374 within the European Community FP7.
MF also thanks the Service de Physique Th\'eorique of the Universit\'e Libre de Bruxelles for hospitality.
The work of TH is supported by the FNRS-FRS, the IISN and the Belgian Science Policy (IAP VI-11). TH thanks J.R.Espinosa and the IFAE at Barcelona for hospitality.
The work of EM has been partly supported by CICYT-FEDER-FPA2008-01430 and by 2009SGR894.

\bibliographystyle{unsrt}

\end{document}